\documentclass[12pt,table]{amsart}
\def\preambleEN{}

\usepackage[utf8]{inputenc}
\usepackage{todonotes}
\ifx\preambleEN\undefined
\usepackage[french]{babel}
\else
\usepackage[english]{babel}
\fi
\usepackage{array} 
\usepackage{xspace} 
\usepackage{xcolor} 

\usepackage{amsmath,amsthm,amsfonts,amssymb,mathrsfs} 
\usepackage{mathtools}
\usepackage{faktor} 
\usepackage{stmaryrd} 

\ifx\preambleEN\undefined
\theoremstyle{plain}
\newtheorem{thm}{Théorème}[section]
\newtheorem*{thm*}{Théorème}

\newtheorem{prop}[thm]{Proposition}
\newtheorem*{prop*}{Proposition}

\newtheorem{lemma}[thm]{Lemme} 
\newtheorem*{lemma*}{Lemme}

\newtheorem{coro}[thm]{Corollaire}
\newtheorem*{coro*}{Corollaire}

\theoremstyle{definition}
\newtheorem{defi}[thm]{Définition}
\newtheorem*{defi*}{Définition}

\theoremstyle{remark}
\newtheorem{rmk}[thm]{Remarque}
\newtheorem*{rmk*}{Remarque}

\newtheorem{example}[thm]{Exemple} 
\newtheorem*{example*}{Exemple}

\else

\theoremstyle{plain}
\newtheorem{thm}{Theorem}[section]
\newtheorem*{thm*}{Theorem}

\newtheorem{prop}[thm]{Proposition}
\newtheorem*{prop*}{Proposition}

\newtheorem{lemma}[thm]{Lemma} 
\newtheorem*{lemma*}{Lemma}

\newtheorem{coro}[thm]{Corollary}
\newtheorem*{coro*}{Corollary}

\theoremstyle{definition}
\newtheorem{defi}[thm]{Definition}
\newtheorem*{defi*}{Definition}

\theoremstyle{remark}
\newtheorem{rmk}[thm]{Remark}
\newtheorem*{rmk*}{Remark}

\newtheorem{example}[thm]{Example} 
\newtheorem*{example*}{Example}

 \addto\captionsenglish{
 }
\fi

\makeatletter
\newcommand\RedeclareMathOperator{%
  \@ifstar{\def\rmo@s{m}\rmo@redeclare}{\def\rmo@s{o}\rmo@redeclare}%
}
\newcommand\rmo@redeclare[2]{%
  \begingroup \escapechar\m@ne\xdef\@gtempa{{\string#1}}\endgroup
  \expandafter\@ifundefined\@gtempa
     {\@latex@error{\noexpand#1undefined}\@ehc}%
     \relax
  \expandafter\rmo@declmathop\rmo@s{#1}{#2}}
\newcommand\rmo@declmathop[3]{%
  \DeclareRobustCommand{#2}{\qopname\newmcodes@#1{#3}}%
}
\@onlypreamble\RedeclareMathOperator
\makeatother

\usepackage{hyperref}

\usepackage{tikz}
\usetikzlibrary{automata,positioning,arrows.meta,cd,babel,shapes}
\usepackage{graphicx} 

\makeatletter
\newcommand{\amslabel}[1]{%
  \Hy@MakeCurrentHref{\@currenvir.\the\Hy@linkcounter}
  \Hy@raisedlink{\hyper@anchorstart{\@currentHref}\hyper@anchorend}
  \label{#1}%
}
\makeatother

\renewcommand{\phi}{\varphi}
\renewcommand{\epsilon}{\varepsilon}

\newcommand{\ens}[2]{\left\{#1\middle|#2\right\}} 
\newcommand{\nN}{\mathbb{N}} 
\DeclareMathOperator{\zZ}{\mathbb{Z}}
\newcommand{\point}[1]{\left\{#1\right\}} 
\newcommand\restr[2]{{
  \left.\kern-\nulldelimiterspace 
  #1 
  \vphantom{\big|} 
  \right|_{#2} 
  }}

\DeclareMathOperator{\Grp}{Grp}
\newcommand{\permut}{\mathfrak{S}}

\newcommand{\noms}{\mathbb{A}}

\DeclareMathOperator{\Nom}{Nom}

\DeclareMathOperator{\Mon}{Mon}
\DeclareMathOperator{\stab}{stab}

\DeclareMathOperator{\op}{^{op}}

\DeclareMathOperator{\Aut}{Aut}

\DeclareMathOperator{\mor}{Mor}

\DeclareMathOperator{\id}{id}

\newcommand{\cC}{\mathcal{C}}
\newcommand{\dD}{\mathcal{D}}

\newcommand{\Cc}{\mathscr{C}}
\newcommand{\Dd}{\mathscr{D}}

\RedeclareMathOperator{\Im}{Im}
\newcommand{\xrightarrowdbl}[1]{\xrightarrow{}\mathrel{\mkern-14mu}\xrightarrow{#1}}

\DeclareMathOperator{\colim}{colim}

\DeclareMathOperator{\comp}{comp} 
\DeclareMathOperator{\Eq}{Eq}

\usepackage{dsfont} 
\newcommand{\terminal}{\mathds{1}}
\newcommand{\initial}{\emptyset}
\DeclareMathOperator{\Min}{Min}

\DeclareMathOperator{\Lan}{Lan}
\DeclareMathOperator{\Ran}{Ran}

\DeclareMathOperator{\Set}{Set}

\DeclareMathOperator{\sets}{Set} 
\newcommand{\eE}{\mathcal{E}}
\newcommand{\fF}{\mathcal{F}}
\newcommand{\gG}{\mathcal{G}}

\newcommand{\classif}{\Omega}

\newcommand{\singleton}[1]{\{\cdot\}_{#1}} 
\DeclareMathOperator{\Rel}{Rel} 
\DeclareMathOperator{\Par}{Par} 
\DeclareMathOperator{\Psh}{Psh} 
\DeclareMathOperator{\Sh}{Sh} 
\newcommand{\khi}{\chi}

\DeclareMathOperator{\ev}{ev} 

\DeclareMathOperator{\Geom}{Geom}
\DeclareMathOperator{\espcl}{\mathbb{B}}
\newcommand{\nom}[1]{\left\ulcorner #1\right\urcorner}

\DeclareFontFamily{U}{min}{}
\DeclareFontShape{U}{min}{m}{n}{<-> udmj30}{}

\DeclareMathOperator{\objin}{in}
\DeclareMathOperator{\objout}{out}
\DeclareMathOperator{\objstates}{st}
\DeclareMathOperator{\iword}{\mathcal{I}}
\DeclareMathOperator{\oword}{\mathcal{O}}

\DeclareMathOperator{\Cat}{\mathcal{C}at}

\newcommand{\aA}{\mathcal{A}}
\newcommand{\bB}{\mathcal{B}}
\newcommand{\uaA}{\underline{\mathcal{A}}}
\newcommand{\ubB}{\underline{\mathcal{B}}}
\DeclareMathOperator{\autocat}{Auto}

\DeclareMathOperator{\Syn}{Syn}

\newcommand{\kleene}[1]{{#1}^\ast}

\newcommand{\vV}{\mathcal{V}}
\DeclareMathOperator{\Quiv}{Quiv}

\usepackage[left=2cm,right=2cm,top=2cm,bottom=2cm]{geometry}
\usepackage{quiver}

\title{Automata in toposes, and general Myhill-Nerode theorems}
\thanks{This project has been partially funded by the European Research Council (ERC) under the European Union's Horizon 2020 research and innovation program (grant agreement No.670624).}
\author{Victor Iwaniack}
\address{Universit\'e C\^ote d'Azur, Laboratoire J. A. Dieudonn\'e}
\email{victor.iwaniack@unice.fr}
\subjclass[2020]{Primary 18B20, 18B25; Secondary 68Q45, 18D20}

\begin{document}

\begin{abstract}
  We extend the functorial approach to automata by Colcombet and Petri\c{s}an \cite{colcombetAutomataMinimizationFunctorial2020} from the category of sets to any elementary topos with natural number object and establish general Myhill-Nerode theorems in our setting. As a special case we recover the result of Boja\'nczyk, Klin and Lasota \cite{bojanczykAutomataTheoryNominal2014} for orbit-finite nominal automata by considering automata in the Myhill-Schanuel topos of nominal sets.
\end{abstract}

\maketitle

\section*{Introduction}

Our purpose here is to extend the categorical approach to automata theory by Colcombet and Petri\c{s}an \cite{colcombetAutomataMinimizationFunctorial2020} to more general contexts than those considered by the authors, namely to automata in an arbitrary elementary topos with a natural number object. One of the notions crucial to the definition of an automaton in such a general context is ``finiteness'' and we will consider two different notions of finiteness which are well-established in topos theory: dK-finiteness (decidable Kuratowski finiteness) and decomposition-finiteness, both reducing to the classical notion of a finite set in the topos of sets. For both notions of finiteness we get a corresponding Myhill-Nerode Theorem characterizing languages with a Nerode congruence of ``finite type'' as those recognized by ``finite type'' automata. The key property beneath these general Myhill-Nerode Theorems is the stability of ``finite'' objects under taking subquotients.

In Section \ref{sec:toposes} we recall basic facts and definitions about toposes and natural number objects, in Section \ref{sec:notions-finite} we discuss Kuratowski and decomposition finiteness, in Section \ref{sec:autom-topos-myhill-nerode} we enrich the functorial approach of Colcombet and Petri\c{s}an and use it to deduce Myhill-Nerode type theorems, and in the last sections we explore automata theory in specific toposes: toposes of $G$-sets for a discrete group $G$ (Subsection \ref{sec:equiv-autom}), toposes of sheaves over a topological space (Subsection \ref{sec:etales-autom}), and finally the Myhill-Schanuel topos of nominal sets (Subsection \ref{sec:nom-autom}).

\subsection*{Acknowledgements}
\label{sec:acknoledgements}

The author would like to thank Clemens Berger for the discussions and the help for the redaction of this article.

\subsection*{Notations}
\label{sec:notations}

We use the \emph{diagrammatical order} for composition: if $f:A\xrightarrow{}B$ and $g:B\xrightarrow{}C$
are morphisms of some category, $fg$ is their composition.

When we consider a category $\mathscr{C}$ enriched over a monoidal category $\vV=(\vV_0,\otimes,I,\alpha,\lambda,\rho)$,
we denote $\Cc_0$ the \emph{underlying
category} with same class of objects as $\Cc$ and with hom-class $\Cc_0(a,b)=\vV(I,\Cc(a,b))$ for any two objects $a$ and $b$.

By ``factorization system'' we will always mean ``orthogonal factorization system'' unless stated otherwise.

If $\Dd\xleftarrow{L}\Cc\xleftarrow{R}\Dd$ is a diagram of functors, then we denote $L\dashv R$ the fact that $L$ is left adjoint to $R$ and will usually denote the unit $\eta:\id_\Cc\xRightarrow{}LR$ and the counit $\epsilon:RL\xRightarrow{}\id_\Dd$.
\section{Toposes}
\label{sec:toposes}

\subsection{Elementary toposes and geometric morphisms}
\label{sec:elem-topos}

\begin{defi}
  An \emph{elementary topos} is a category $\eE$ with
  \begin{enumerate}
  \item finite limits;
  \item exponentials, i.e.\ for each object $B$ of $\eE$, the endofunctor $(-)\times B : \eE \xrightarrow{} \eE$ has a right adjoint denoted $(-)^{B}$ or $[B,-]$;
  \item a subobject classifier i.e.\ an object $\classif$ of $\eE$ equipped with a morphism
  $\terminal\xrightarrow{\top}\classif$ such that for each object $A$ and subobject $S\xhookrightarrow{}A$, there exists a unique morphism $\khi_S:A\xrightarrow{}\classif$ called the \emph{characteristic map} such that the following diagram
    \begin{center}
      \begin{tikzcd}
        {S} & {\terminal} \\
        {A} & {\classif}
        \arrow[""', hook,from=1-1, to=2-1]
        \arrow["{!}", from=1-1, to=1-2]
        \arrow["{\top}", from=1-2, to=2-2]
        \arrow["{\khi_S}"', from=2-1, to=2-2]
        \arrow["\lrcorner"{anchor=center, pos=0.125}, draw=none, from=1-1, to=2-2]
      \end{tikzcd}
    \end{center}
    is a pullback.
  \end{enumerate}
\end{defi}

\begin{rmk}
  The existence of a subobject classifier amounts to the existence of a right adjoint
    $\mathcal{P}:\Rel(\eE)\xrightarrow{}\eE$ to the inclusion functor of $\eE$ into
    $\Rel(\eE)$, sending a morphism $A\xrightarrow{f}B$ to the \emph{functional} relation $A\xleftarrow{\id_A}A\xrightarrow{f}B$.
    In that case, $\classif=\mathcal{P}(\terminal)$ and $\top$ is the unit at $\terminal$. For an object $A$, $\mathcal{P}(A)$ is
    the \emph{power object} of $A$, and by adjunction global elements $\terminal\xrightarrow{}\mathcal{P}(A)$ correspond bijectively
    to relations $S \subset \terminal\times A$, namely subobjects of $A$. If $\eE$ has exponentials, then $\mathcal{P}(A)=\classif^A$.
\end{rmk}

\begin{prop}
  An elementary topos is a pretopos (i.e.\ a category that is simultaneously Barr-exact and extensive) as well as a Heyting category.
\end{prop}

\begin{defi}
  An object $A$ of an elementary topos $\eE$ is \emph{decidable} if the diagonal $A\xhookrightarrow{(\id_A,\id_A)}A\times A$ is a \emph{complemented subobject}, i.e.\ if there exists a subobject $C\xhookrightarrow{}A\times A$ such that
  \begin{center}
    \begin{tikzcd}
      \initial & A \\
      C & A\times A
      \arrow[from=1-1,to=2-1]
      \arrow[from=1-1,to=1-2]
      \arrow[hook,from=2-1,to=2-2]
      \arrow[hook,from=1-2,to=2-2]
    \end{tikzcd}
  \end{center}
  is a pullback (which means the intersection $A\cap C=\emptyset$ is initial) as well as a pushout diagram (which means the union $A\cup C= A\times A$ is the whole of $A\times A$).
\end{defi}

We denote $\bot:\terminal\xrightarrow{}\classif$ the characteristic map of the subobject $\initial\xhookrightarrow{}\terminal$.

Decidable objects will be of interest when we will talk about finiteness conditions in a topos. We recall the characterisation of Boolean toposes from Acu\~na-Ortega and Linton \cite[Observation 2.6]{acuna-ortegaFinitenessDecidability1979}
\begin{prop}
  Let $\eE$ be an elementary topos. If one, hence all of the following equivalent assertions is true, then $\eE$ is said to be \emph{Boolean}:
  \begin{enumerate}
    \item $\terminal+\terminal\xrightarrow{(\top,\bot)}\classif$ is an isomorphism
    \item each subobject is complemented
    \item each object is decidable
    \item $\classif$ is decidable
  \end{enumerate}
\end{prop}

\begin{defi}
  Let $\eE$ and $\fF$ be elementary toposes.

  A geometric morphism $f$ between $\eE$ and $\fF$ is an adjunction $(f^\ast \dashv f_\ast)$ such that the left adjoint
    $f^\ast:\fF\xrightarrow{}\eE$ preserves finite limits. The left adjoint is called the \emph{inverse image functor} of $f$, and the
    right one the \emph{direct image functor}.
  \end{defi}

  Because elementary toposes enjoy the solution set condition of the Freyd's adjoint functor theorem, a geometric morphism $f:\eE\xrightarrow{}\fF$ is entirely determined by a left exact, cocontinuous functor $f^\ast:\fF\xrightarrow{}\eE$.

\begin{defi}
  A geometric morphism $f:\eE\xrightarrow{}\fF$ between elementary toposes is said to be 
  \begin{itemize}
  \item \emph{essential} if the left adjoint $f^\ast:\fF\xrightarrow{}\eE$ also has a left adjoint
  \item \emph{atomic} if $f^\ast$ is a \emph{logical functor}, i.e.\ it preserves finite limits (automatic), exponentials and the subobject classifier
  \end{itemize}
\end{defi}

\subsection{Grothendieck toposes}
\label{sec:grothendieck-toposes}

\begin{defi}
  A category $\eE$ is a Grothendieck topos if it is equivalent to a category of \emph{sheaves} $\Sh(\Cc,J)$ on a \emph{small site} $(\Cc,J)$ (meaning $\Cc$ is a small category).
\end{defi}

\begin{thm}
  A category $\eE$ is a Grothendieck topos if and only if it satisfies one (and hence all) of the following conditions:
  \begin{enumerate}
  \item $\eE$ is a reflective subcategory $i:\eE\xhookrightarrow{}\Psh(\Cc)$
    of a presheaf topos where the reflector $r$ (the left adjoint of the inclusion functor) is left exact
  \item $\eE$ satisfies the Giraud axioms:
    \begin{enumerate}
    \item $\eE$ admits a set of generators
    \item has all finite limits
    \item has all small coproducts, which are disjoints and stable under pullback
    \item every congruence has a quotient, and those quotients are stable under pullback
    \end{enumerate}
  \end{enumerate}
\end{thm}

The following theorem may be called the fundamental theorem of topos theory and we will make use of it without citing it (see Mac Lane and Moerdijk \cite[Section IV.7]{maclaneSheavesGeometryLogic1994} and Artin, Grothendieck and Verdier\cite[Section III.5]{artinTheorieToposCohomologie}):

\begin{thm*}
  \begin{itemize}
  \item If $\eE$ is an elementary topos then so is $\eE/A$ for all objects $A$.
  \item If $\eE$ is a Grothendieck topos then so is $\eE/A$ for all objects $A$.
  \end{itemize}
\end{thm*}

\begin{prop}
  A Grothendieck topos $\eE$ is an elementary topos. It is equipped with a \emph{global section geometric morphism} $\gamma:\eE\xrightarrow{}\Set$ and it is the only such geometric morphism (up to natural isomorphism) and it is defined by
  $$\gamma^*(S)=\sum_{s\in S}\terminal$$
  $$\gamma_*(\fF)=\eE(\terminal,\fF)$$
\end{prop}

\begin{defi}
  In an elementary topos $\eE$, an object $A$ is said to be \emph{connected} if it has exactly two complemented subobjects, namely $\initial$ and $A$ itself.
\end{defi}

\begin{rmk}
  Equivalently, $A$ is connected iff $\eE(A,-)$ preserves binary coproducts. Note that in particular, $\initial$ is not connected.
\end{rmk}

\begin{prop}
  A Grothendieck topos $\eE$ is locally connected (i.e.\ the morphism $\gamma:\eE\xrightarrow{}\Set$ is essential) iff each object is a coproduct of connected objects.
  A Grothendieck topos $\eE$ is atomic (i.e.\ the morphism $\gamma:\eE\xrightarrow{}\Set$ is atomic) iff $\eE$ is Boolean and locally connected.

  In both cases the inverse image of the global section morphism admits by definition a left adjoint we denote $\pi_0$, such that $\pi_0(A)$ is the set of connected components of an object $A$.
\end{prop}

\begin{proof}
  See Berger and Iwaniack \cite[Lemma 2.2]{bergerProfiniteFundamentalGroup2023}
\end{proof}

\begin{defi}
  Let $\eE$ be a Grothendieck topos with global section morphism $\gamma$, and $A$ an object.
  \begin{itemize}
  \item We say $A$ is \emph{constant} if it is isomorphic to $\gamma^\ast(S)$ for some (small) set $S$.
  \item We say $A$ is \emph{locally constant} if there exists a (small) family of objects $(U_i)_i$ that \emph{covers} $\terminal$ i.e.\ $\sum_{i\in I}U_i\xrightarrowdbl{}\terminal$, such that for all $i\in I$, $U_i\times A$ is a constant object of the Grothendieck topos $\eE/U_i$.
  \item We denote $LC(\eE)$ the full subcategory of $\eE$ spanned by locally constant objects of $\eE$.
  \item We denote $SLC(\eE)$ the full subcategory of $\eE$ spanned by (small) coproducts of locally constant objects of $\eE$.
  \end{itemize}
\end{defi}

The following theorem has first been proved by Leroy \cite[Theorem 2.4]{leroyGroupoideFondamentalTheoreme1979}. The thus arising Grothendieck toposes are often called \emph{Galois toposes}.

\begin{thm}\label{thm:leroy}
  Let $\eE$ be a locally connected Grothendieck topos. Then $SLC(\eE)$ is a locally connected Grothendieck topos generated by its locally constant objects.
\end{thm}

\subsection{Natural number object and monoid objects}
\label{sec:natur-numb-object}

\begin{defi}
  Let $\eE$ be an elementary topos. We say that $\eE$ \emph{has a natural number object} if and only if it has a triple $(\nN_\eE,z_\eE:\terminal\xrightarrow{}\nN_\eE,s_\eE:\nN_\eE\xrightarrow{}\nN_\eE)$ with the following universal property: given any other triple $(A,x_0:\terminal\xrightarrow{}A,f:A\xrightarrow{}A)$, there exists a unique $x_\bullet:\nN_\eE\xrightarrow{}A$
  such that the following diagram commute:
  \begin{center}
    \begin{tikzcd}[ampersand replacement=\&]
        \terminal \& \nN_\eE \& \nN_\eE \\
        \& A \& A
        \arrow["{x_0}", from=1-1, to=2-2]
        \arrow["z_\eE", from=1-1, to=1-2]
        \arrow["s_\eE", from=1-2, to=1-3]
        \arrow["f", from=2-2, to=2-3]
        \arrow["{x_\bullet}", from=1-2, to=2-2]
        \arrow["{x_\bullet}", from=1-3, to=2-3]
      \end{tikzcd}
  \end{center}
\end{defi}
The idea is that we can recursively define sequences $x_\bullet:\nN_\eE\xrightarrow{}A$ by providing the first term $x_0$
and the function $f$ such that ``$x_{n+1}=f(x_n)$''

With this property we can define the predecessor morphism $p_\eE$ so that ``$p(0)=0$, $p(n+1)=n$'', but because we have to remember $n$, we define ``$x_\bullet(0)=(0,0)$, $x_\bullet(n+1)=(x_{\bullet,2}(n),x_{\bullet,2}(n)+1)$''

\begin{center}
  \begin{tikzcd}[ampersand replacement=\&]
    \terminal \& \nN_\eE \& [+30pt] \nN_\eE \\
    \& \nN_\eE\times \nN_\eE \& \nN_\eE\times \nN_\eE
    \arrow["{(z_\eE,z_\eE)}"', from=1-1, to=2-2]
    \arrow["z_\eE", from=1-1, to=1-2]
    \arrow["s_\eE", from=1-2, to=1-3]
    \arrow["{x_\bullet}", from=1-2, to=2-2]
    \arrow["{x_\bullet}", from=1-3, to=2-3]
    \arrow["\pi_2\Delta_{\nN_\eE}(\nN_\eE\times s_\eE)", from=2-2, to=2-3]
  \end{tikzcd}
\end{center}
and thus let $p_\eE := x_\bullet\pi_2$, and the truncated subtraction $t_\eE:\nN_\eE\times\nN_\eE\xrightarrow{}\nN_\eE$ defined by its adjunct ${t_\eE}^\dashv$, so that
``$t(m)(0)=m-0=m$, $t(m)(n+1)=m-(n+1)=(m-n)-1=p(t(m)(n))$'':

\begin{center}
  \begin{tikzcd}[ampersand replacement=\&]
    \terminal \& \nN_\eE \& \nN_\eE \\
    \& {\nN_\eE^{\nN_\eE}} \& {\nN_\eE^{\nN_\eE}}
    \arrow["{\id_{\nN_\eE}}"', from=1-1, to=2-2]
    \arrow["z_\eE", from=1-1, to=1-2]
    \arrow["s_\eE", from=1-2, to=1-3]
    \arrow["{{t_\eE}^\dashv}", from=1-2, to=2-2]
    \arrow["{{t_\eE}^\dashv}", from=1-3, to=2-3]
    \arrow["{{p_\eE}^{\nN_\eE}}", from=2-2, to=2-3]
  \end{tikzcd}
\end{center}
The truncated subtraction allows us to define the object of order pairs of natural numbers $(i,j),i\leq j$ by the pullback
\begin{center}
  \begin{tikzcd}[ampersand replacement=\&]
    {\nN_{\eE,\leq}} \& \terminal \\
    \nN_\eE\times\nN_\eE \& \nN_\eE
    \arrow["z_\eE", hook, from=1-2, to=2-2]
    \arrow["t_\eE", from=2-1, to=2-2]
    \arrow["{!}", from=1-1, to=1-2]
    \arrow[hook, from=1-1, to=2-1]
    \arrow["\lrcorner"{anchor=center, pos=0.125}, draw=none, from=1-1, to=2-2]
  \end{tikzcd}
\end{center}

This relation on $\nN_\eE$ is useful to define the notion of internal finite cardinals.

\begin{defi}
  Let $\eE$ be a topos with a natural number object. An object of $\eE$ is a \emph{finite cardinal} if it is isomorphic
  to an object $[n]$ obtained as a pullback
  \begin{center}
    \begin{tikzcd}[ampersand replacement=\&]
      {[n]} \&\& \terminal \\
      {\nN_{\eE,\leq}} \& \nN_\eE\times\nN_\eE \& \nN_\eE
      \arrow["n", hook, from=1-3, to=2-3]
      \arrow["{!}", from=1-1, to=1-3]
      \arrow[hook, from=1-1, to=2-1]
      \arrow["{\pi_2\times s}", from=2-2, to=2-3]
      \arrow[hook, from=2-1, to=2-2]
      \arrow["\lrcorner"{anchor=center, pos=0.125}, draw=none, from=1-1, to=2-2]
    \end{tikzcd}
  \end{center}
  for some global element $n:\terminal\xrightarrow{}\nN_\eE$.
\end{defi}

\begin{prop}\label{prop:arith-transf-inv-img}
  Let $\eE\xrightarrow{f}\fF$ be a geometric morphism between elementary toposes. Suppose $\fF$ admits a natural number object $\nN_\fF$.
  \begin{enumerate}
  \item $(f^*(\nN_\fF),f^*(z_\fF),f^*(s_\fF))$ is a natural number object in $\eE$ (well defined because $f^*$ preserves the terminal object).
  \item $p_\eE=f^*(p_\fF)$ so that $t_\eE=f^*(t_\fF)$ and $\nN_{\leq,\eE}=f^*(\nN_{\leq,\fF})$.
  \end{enumerate}
\end{prop}

\begin{proof}
  \begin{enumerate}
  \item First denote $(0,1)(\eE)$ the category of absolute $(0,1)$-algebra objects in $\eE$. The adjunction $f^* \dashv f_* : \fF\xrightarrow{}\eE$ can be lifted to $\overline{f^*} \dashv \overline{f_*} : (0,1)(\fF)\xrightarrow{}(0,1)(\eE)$ by letting $f^*$ and $f_*$ acting componentwise on algebras $(A,\terminal\xrightarrow{e}A,A\xrightarrow{u}A)$ provided $f^*$ and $f_*$ both preserve the terminal object. It is a lifting because if we denote $U_\eE$ and $U_\fF$ the forgetful functors from the categories of $(0,1)$-algebras to their respective topos of definition, then we see that $(U_\fF,U_\eE)$ is a morphism of adjunctions so that $\overline{f^*}U_\eE=U_\fF f^*$.

    Then $\overline{f^*}$ being a left adjoint it preserves the natural number object as it is the initial object of $(0,1)(\fF)$.
  \item Because $f^*$ preserves finite products, commutation of the diagram corresponding to $p_\fF$ makes the diagram
    \begin{center}
      \begin{tikzcd}[ampersand replacement=\&]
        \terminal \& f^*(\nN_\fF) \& [+90pt] f^*(\nN_\fF) \\ [+15pt]
        \& f^*(\nN_\fF)\times f^*(\nN_\fF) \& f^*(\nN_\fF)\times f^*(\nN_\fF)
        \arrow["{(f^*(z_\fF),f^*(z_\fF))}"', from=1-1, to=2-2]
        \arrow["f^*(z_\fF)", from=1-1, to=1-2]
        \arrow["f^*(s_\fF)", from=1-2, to=1-3]
        \arrow["{f^*(x_\bullet)}", from=1-2, to=2-2]
        \arrow["{f^*(x_\bullet)}", from=1-3, to=2-3]
        \arrow["\pi_2\Delta_{f^*(\nN_\fF)}(f^*(\nN_\fF)\times f^*(s_\fF))", from=2-2, to=2-3]
      \end{tikzcd}
    \end{center}
    and then $f^*(x_\bullet)\pi_2=f^*(x_\bullet\pi_2)=f^*(p_\fF)$.
    
    Commutation of the diagram defining ${t_\fF}^\dashv$ implies the commutation of the adjoint diagram
    \begin{center}
      \begin{tikzcd}[ampersand replacement=\&]
        \nN_\fF\times \terminal \& \nN_\fF\times\nN_\fF \& \nN_\fF\times\nN_\fF \\
        \& {\nN_\fF} \& {\nN_\fF}
        \arrow["\sim"', from=1-1, to=2-2]
        \arrow["\nN_\fF\times z_\fF", from=1-1, to=1-2]
        \arrow["\nN_\fF\times s_\fF", from=1-2, to=1-3]
        \arrow["{t_\fF}", from=1-2, to=2-2]
        \arrow["{t_\fF}", from=1-3, to=2-3]
        \arrow["{p_\fF}", from=2-2, to=2-3]
      \end{tikzcd}
    \end{center}
    and then by taking the image of this diagram by $f^*$ and taking the adjoint we have a commuting diagram showing that $t_\eE=f^*(t_\fF)$.
    
    Finally, the pullback of $\nN_{\fF,\leq}$ is preserved by $f^*$ so that $\nN_{\eE,\leq}=f^*(\nN_{\fF,\leq})$.
  \end{enumerate}
\end{proof}

\begin{coro}
  If $\eE$ is a Grothendieck topos, then $\nN_\eE=\gamma^*(\nN)$ and $\nN_{\eE,\leq}=\gamma^*(\nN_\leq)$.
\end{coro}

The reason why the natural number object is of interest here is that we need the existence of free (internal) monoids in order to define languages, and
we will see that in an elementary topos, the existence of free monoids is equivalent to the existence of a natural number object.

\begin{defi}
  Let $\eE$ be a category with finite limits. An \emph{internal monoid} or a \emph{monoid object} of $\eE$
  is a tuple $(M,\eta:\terminal\xrightarrow{}M,\mu:M\times M\xrightarrow{}M)$ satisfying the following commuting
  diagrams:
  \begin{description}
  \item[associativity]
    \begin{center}
      \begin{tikzcd}[ampersand replacement=\&]
        {M\times M\times M} \& {M\times M} \\
        {M\times M} \& M
        \arrow["{\mu\times M}", from=1-1, to=2-1]
        \arrow["{M\times \mu}"', from=1-1, to=1-2]
        \arrow["\mu"', from=1-2, to=2-2]
        \arrow["\mu", from=2-1, to=2-2]
      \end{tikzcd}
    \end{center}
  \item[unitality]
    \begin{center}
      \begin{tikzcd}[ampersand replacement=\&]
        M \& {M\times M} \\
        {M\times M} \& M
        \arrow["{\eta\times M}", from=1-1, to=2-1]
        \arrow["M\times\eta", from=1-1, to=1-2]
        \arrow["\mu"', from=1-2, to=2-2]
        \arrow["\mu", from=2-1, to=2-2]
        \arrow["\id_M", from=1-1, to=2-2]
      \end{tikzcd}
    \end{center}
  \end{description}
  The category $\Mon(\eE)$ has objects the internal monoids of $\eE$, and morphisms the
  \emph{homomorphisms of internal monoids}: $f:(M,\mu,\eta)\xrightarrow{}(N,\nu,\theta)$ is a homomorphism
  if $f:M\xrightarrow{}N$ is a morphism of $\eE$ such that
  \begin{center}
    \begin{tikzcd}[ampersand replacement=\&]
      {M\times M} \& {N\times N} \\
      M \& N
      \arrow["\mu", from=1-1, to=2-1]
      \arrow["{f\times f}", from=1-1, to=1-2]
      \arrow["f", from=2-1, to=2-2]
      \arrow["\nu"', from=1-2, to=2-2]
    \end{tikzcd}
    and
    \begin{tikzcd}[ampersand replacement=\&]
      \terminal \\
      M \& N
      \arrow["\eta", from=1-1, to=2-1]
      \arrow["\theta", from=1-1, to=2-2]
      \arrow["f", from=2-1, to=2-2]
    \end{tikzcd}
  \end{center}
  commute.

  The category $\Mon(\eE)$ comes with a canonical faithful \emph{forgetful functor}
  $$\mathcal{U}:\Mon(\eE)\xrightarrow{}\eE$$
  sending an arrow $f:(M,\mu,\eta)\xrightarrow{}(N,\nu,\theta)$ to $f:M\xrightarrow{}N$.
  
  Thus, we say $\eE$ \emph{has free (internal) monoids} if $\mathcal{U}$ admits a left adjoint then denoted $(-)^\ast$,
  sending an object $\Sigma$ to the monoid $(\kleene{\Sigma},m_\Sigma,\epsilon_\Sigma)$.
\end{defi}

\begin{rmk}
  For any object $A$ of an elementary topos $\eE$, the exponential $A^A$ is canonically endowed with a monoid structure, with multiplication
  $\comp_{A,A,A}$ and neutral element the adjunct $\terminal\xrightarrow{}A^A$ of the identity of $A$,
  because it is an endormorphism object of $\eE$ as an $\eE$-category.
\end{rmk}

The following proposition of Johnstone \cite[Proposition 5.3.3]{johnstoneSketchesElephant2002} explains why we need a natural number object in the context of language theory.

\begin{prop}\label{prop:elem-top-has-nno-iff-free-mon}
  A elementary topos has a natural number object if and only if it has free internal monoids.
\end{prop}

\begin{prop}
  A Grothendieck topos the free internal monoid on an object $A$ has underlying object $\sum_{n\in \nN}A^n$. In particular the natural number object is $\gamma^\ast(\nN)$. More generally, if the topos is not Grothendieck but the coproduct exists, then it defines the free monoid generated by $A$.
\end{prop}

\begin{proof}
  The assertion for the natural number object can be seen as a consequence of Proposition \ref{prop:arith-transf-inv-img}, but the general assertion can be verified by returning to the definition of the free monoid.
\end{proof}

\begin{rmk}
  One of the key examples of the article are automata in the Grothendieck topos of continuous $G$-sets, denoted $\espcl G$, where $G$ is a topological group acting continuously on sets, viewed as discrete topological spaces, see Mac Lane and Moerdijk \cite{maclaneSheavesGeometryLogic1994} for a proof that these form indeed a Grothendieck topos.
  In particular, automata in $\espcl\Aut(\nN)$, where $\Aut(\nN)$ is the topological group of symmetries of $\nN$ (the topology
  being inherited from the product topology of $\prod_{n\in\nN}\nN$, where $\nN$ has the discrete topology) are called nominal automata, the Grothendieck topos $\espcl\Aut(\nN)$ being called the topos of nominal sets. Those automata are as expressive as finite-memory automata, see Boja\'nczyk, Klin and Lasota \cite[Theorems 6.4 and 6.6]{bojanczykAutomataTheoryNominal2014} for a proof of the equivalence. We usually consider that the group acting is the group of symmetries of a countably infinite
  set $\noms$ of ``atoms'' or ``names'' instead of $\nN$ itself.

  Given a (topological or not) group $G$, and a continuous $G$-set $A$, the internal free monoid on $A$ is the set $\kleene{A}$ with the action defined as follows:
for any word $w=a_1a_2\cdots a_n$ on $A$, such that $a_i\in A$, and $g\in G$, $w\cdot g := (a_1\cdot g)(a_2\cdot g)\cdots(a_n\cdot g)$ because in $\espcl G$, finite limits and small colimits are computed pointwise.
\end{rmk}

\section{Notions of finiteness}
\label{sec:notions-finite}

\begin{defi}
  Let $A$ be an object of an elementary topos $\eE$.
  The sub-monoid of $(\classif^A,\vee,\initial)$ generated by the singleton subobject $\singleton{A}:A\xhookrightarrow{}\classif^A$ (adjoint to the characteristic morphism of the diagonal $A\xhookrightarrow{(\id_A,\id_A)}A\times A$) is denoted $K(A)$ and called the object of \emph{Kuratowski-finite subobjects of $A$}.
\end{defi}

\begin{defi}
  Let $\eE$ be an elementary topos.
  \begin{enumerate}
  \item An object is called \emph{decomposition-finite} if it is a finite coproduct of connected subobjects.
  \item An object $A$ is \emph{Kuratowski-finite} or \emph{K-finite} if the global element $\terminal\xrightarrow{}\classif^A$ corresponding to $A\subset A$ factors through $K(A)\subset \classif^A$.
  \item We say the object is \emph{decidable Kuratowski finite}, we abbreviate dK-finite, if it is both decidable and Kuratowski finite.
  \item Let $P$ be a non-empty class of points of $\eE$ (usually, $\eE$ will be a Grothendieck topos with enough points and $P$ a sufficient set of points to decide isomorphy), an object $A$ is \emph{$P$-stalkwise finite} if for all point $x$ in $P$, the stalk of $A$ at $x$ is finite i.e.\ $x^\ast(A)$ is a finite set.
  \end{enumerate}
  \end{defi}

\begin{prop}\label{prop:stlk-fin-stable-subquo}
  stalkwise finiteness is stable under subquotients.
\end{prop}

\begin{proof}
  This is simply due to the fact that the functor taking a sheaf to its stalk at $x$ is the inverse image of a geometric morphism and therefore preserves both monomorphisms and epimorphisms.
\end{proof}

Details about K-finiteness can be found at Johnstone \cite[Subsection D5.4]{johnstoneSketchesElephant2002}, in particular Theorem 5.4.13.
\begin{prop}\label{prop:fin-conds-stable-subquo}
  Both dK- and decomposition-finiteness are stable under complemented subobjects and decidable quotients.
  
  The decomposition-finiteness (for any atomic Grothendieck topos $\eE$) and K-finiteness (for any Boolean topos $\eE$) are stable under subquotients.
\end{prop}

\begin{proof}
  In the order:
  \begin{itemize}
  \item According to Johnstone \cite[Lemma A2.4.8]{johnstoneSketchesElephant2002}, $\pi_0$ preserves monomorphisms as it is left adjoint to a logical functor between toposes. Moreover, as a left adjoint, it preserves epimorphisms as well.
  \item Every subobject in a Boolean topos is complemented, so by Johnstone \cite[Lemma 5.4.4.iv]{johnstoneSketchesElephant2002} a subobject of a K-finite object is K-finite. In any topos, K-finite object are closed under quotients according to Johnstone \cite[Lemma 5.4.4.ii]{johnstoneSketchesElephant2002}.
  \end{itemize}
\end{proof}

\begin{rmk}
  To see why, for decomposition-finiteness, $\pi_0$ might not preserve subquotients if $\eE$ is only locally connected
  but not atomic, consider the topos of sheaves over the circle. It is locally connected because the circle
  is locally connected, and it entails that each étalé space over the circle is a coproduct of connected étalés spaces over
  the circle. Now think about the open two halves of the circle as étalé over the circle. Then it has two connected components,
  is a subobject of the circle (as an étalé space over the circle itself), but the set of connected component of the former
  (with two connected components) is not a subobject of the latter (with one connected component, itself). So in this
  case $\pi_0$ does not preserve monomorphisms. Its does not even preserve finiteness: consider the subétalé space of
  the circle defined as the coproduct of the open first half of the circle, then the next quarter of the circle, then
  the next open eighth of the circle, et cetera. It has countably infinitely many connected component and is a subobject
  of a connected étalé space.
\end{rmk}

\begin{thm}\label{thm:dk-fin-in-groth-topos-charac}
  For any object $A$ of a Grothendieck topos $\eE$, the following assertions are equivalent:
  \begin{enumerate}
  \item $A$ is dK-finite
  \item there exists an epimorphism $!:U\xrightarrowdbl{}\terminal$ such that $U\times A$ is isomorphic to a finite cardinal in the topos $\eE_{/U}$
  \item $A$ is locally finite i.e.\ there exists an epimorphism $\sum_{i\in I}U_i\xrightarrowdbl{}\terminal$ from a coproduct of a family of objects $(U_i)_{i\in I}$ such that, for all $i\in I$, $U_i\times A$ is isomorphic to a finite cardinal in the topos $\eE_{/U_i}$
  \end{enumerate}
\end{thm}

\begin{proof}
  (1) is equivalent to (2) because by Johnstone \cite[Theorem 5.4.13]{johnstoneSketchesElephant2002}, dK-finite objects are exactly decidable objects that are locally a quotient of a finite cardinal, but decidability is a local notion, and according to Johnstone \cite[Corollary 5.2.6]{johnstoneSketchesElephant2002}, decidable quotients of a finite cardinal are finite cardinals.

  The equivalence of (1) and (3) is done in Berger and Iwaniack \cite[Proposition 3.6]{bergerProfiniteFundamentalGroup2023}.
\end{proof}

\section{Automata in toposes and Myhill-Nerode type theorems}
\label{sec:autom-topos-myhill-nerode}

Before enriching the approach of Colcombet and Petri\c{s}an \cite{colcombetAutomataMinimizationFunctorial2020}, we recall their point of view.
Consider a complete deterministic automaton $(Q,i,F,\delta)$ on an alphabet $\Sigma$ (any finite non-empty set), meaning $Q$ is a set of \emph{states}, $i\in Q$ the \emph{initial state}, $F\subset Q$ the set of \emph{final states} and $\delta:Q\times \Sigma\xrightarrow{}Q$ the \emph{transition} function. The transition function gives, by iteration, a right action of the free monoid $\kleene{\Sigma}$ generated by $\Sigma$, on the set $Q$.
In particular we can interpret the action in a functorial (classical) way as a functor $\kleene{\Sigma}\xrightarrow{}\Set$ where the monoid $\kleene{\Sigma}$ is seen as a category with a single object $\objstates$.
Now the initial state $i\in Q$ can be seen as a global element $\terminal\xrightarrow{}Q$, and the
subset $F$ of final states can be represented by its characteristic morphism $\khi_F:Q\xrightarrow{}\classif$ where $\classif$ is the subobject classifier of $\sets$, namely any two-element set of ``truth values''.
All this data can be expressed by a functor $\iword_\Sigma\xrightarrow{} \Set$ with source freely generated by the quiver
\begin{center}
  \begin{tikzcd}
    \objin \arrow[r, "\triangleright"] & \objstates \arrow[r, "\triangleleft"] \arrow["s \in \Sigma"', loop, distance=2em, in=125, out=55] & \objout
  \end{tikzcd}
\end{center}
and the functor corresponding to the complete deterministic automaton $(Q,i,F,\delta)$
\begin{center}
  \begin{tikzcd}
    \terminal \arrow[r, "i"] & Q \arrow[r, "\khi_F"] \arrow["{\delta(-,s), s \in \Sigma}"', loop, distance=2em, in=125, out=55] & \classif
  \end{tikzcd}
\end{center}
sends $(\objin,\objstates,\objout)$ to $(\terminal,Q,\classif)$, $\triangleright$ to the global element corresponding to $i$, $\triangleleft$ to the characteristic function corresponding to $F$, and extends the previous functor $\kleene{\Sigma}\xrightarrow{}\Set$. This correspondence is in fact a bijection.

We can now define what exactly we mean by an automaton in a topos.
\begin{defi}
  Let $\eE$ be an elementary topos with a natural number object, and let $\Sigma$ be an object of this topos, called an \emph{alphabet}.
  \begin{itemize}
  \item A language on $\Sigma$ is any subobject of $\kleene{\Sigma}$
  \item A (deterministic, complete) automaton on the alphabet $\Sigma$, is a quadruple $\aA=(Q,i,F,\delta)$ where
    \begin{itemize}
    \item $Q$ is the \emph{states} object
    \item $i:\terminal\xrightarrow{}Q$ is a global element, the \emph{initial state}, of $Q$
    \item $F$ is the subobject of $Q$ of \emph{final states}, which we identify with its characteristic morphism
      $\khi_F:Q\longrightarrow{}\classif$.
    \item $\delta:Q\times \Sigma\xrightarrow{} Q$ is the \emph{transition morphism}
    \end{itemize}
  \end{itemize}
\end{defi}

There is a notion of a language recognized by an automaton.
To define it, observe that the adjunct, with respect to cartesian closedness of $\eE$, of $\delta:Q\times \Sigma\xrightarrow{} Q$,
we denote $\delta^\dashv:\Sigma\xrightarrow{}Q^Q$, takes values in an internal monoid $Q^Q$.
Because $\kleene{\Sigma}$ is the free internal monoid, $\delta^\dashv$ extends uniquely to $\kleene{\Sigma}$ into
an internal monoids morphism we call $\delta^\ast:\kleene{\Sigma}\xrightarrow{}Q^Q$. This leads to the following definition:
\begin{defi}
  The language recognized by the automaton $\aA=(Q,i,F,\delta)$ o, $\Sigma$ is the subobject $L(\aA)$ of $\kleene{\Sigma}$ corresponding to the following global element: $$\terminal\xrightarrow{i}Q\xrightarrow{\delta^\ast}Q^{\kleene{\Sigma}}\xrightarrow{(\khi_F)^{\kleene{\Sigma}}}\classif^{\kleene{\Sigma}}$$ When $\aA$ recognizes a language $L$, we say $\aA$ is an $L$-automaton.
\end{defi}

\subsection{Languages and automata as enriched functors}

According to Colcombet and Petri\c{s}an \cite{colcombetAutomataMinimizationFunctorial2020}, a complete deterministic automaton may be represented by a functor $\uaA:\iword_\Sigma\xrightarrow{}\sets$ where $\iword_\Sigma$ is the category freely generated by the quiver
\begin{center}
  \begin{tikzcd}
    \objin \arrow[r, "\triangleright"] & \objstates \arrow[r, "\triangleleft"] \arrow["a \in \Sigma"', loop, distance=2em, in=125, out=55] & \objout
\end{tikzcd}
\end{center}
where $\uaA$ sends $(\objin,\objstates,\objout)$ to $(\terminal,Q,\classif)$, and the language recognized by the automaton is encoded by the restriction of $\uaA$ to the full subcategory $\oword_\Sigma$ spanned by the objects $\objin$ and $\objout$.

A crucial point in the construction of $\iword_\Sigma$ is that the endomorphism monoid of $\objstates$ is the free monoid on $\Sigma$.
Now because we would like to consider $\Sigma$ as an object of a topos $\eE$,
the endomorphism monoid should be an internal monoid of $\eE$, therefore we define $\iword_\Sigma$ as a free category enriched in $\eE$ (see \autoref{sec:free-e-categories} for details about existence and construction of free $\eE$-categories).

The definitions we will give in the following subsections are immediate generalisations of the definitions of Colcombet and Petri\c{s}an \cite{colcombetAutomataMinimizationFunctorial2020}, so that we will keep essentially the same terminology.

\subsection{Definitions}

\begin{defi}
  Let $\Sigma$ be an object of a topos $\eE$ admitting a natural number object. The $\eE$-category $\iword_\Sigma$,
  the \emph{$\eE$-category of internal behaviors over the alphabet $\Sigma$}, is the $\eE$-category freely generated by the $\eE$-quiver $Q_\Sigma$ such that
    \begin{itemize}
    \item $(Q_\Sigma)_0=\left\{\objin,\objstates,\objout\right\}$,
    \item $Q_\Sigma(\objin,\objstates)=\terminal=Q_\Sigma(\objstates,\objout)$,
    \item $Q_\Sigma(\objstates,\objstates)=\Sigma$ and
    \item $Q_\Sigma(X,Y)=\initial$ otherwise.
    \end{itemize}
\end{defi}

\begin{rmk}
  Spelled out, $\iword_\Sigma$ is defined by:
  \begin{description}
  \item[Objects] three objects $\objin$, $\objstates$ and $\objout$
  \item[Objects of morphisms] given by the table:
    \begin{center}
      \begin{tabular}{|c|c|c|c|}
        \hline
        $\iword_\Sigma(\downarrow,\rightarrow)$ & $\objin$ & $\objstates$ & $\objout$ \\
        \hline
        $\objin$ & $\terminal$ & $\kleene{\Sigma}$ & $\kleene{\Sigma}$ \\
        \hline
        $\objstates$ & $\initial$ & $\kleene{\Sigma}$ & $\kleene{\Sigma}$ \\
        \hline
        $\objout$ & $\initial$ & $\initial$ & $\terminal$ \\
        \hline
      \end{tabular}
    \end{center}
  \item[Composition morphisms] considering the preceding table, the composition morphisms
    in $\iword_\Sigma$ is of one of the following form:
    \begin{itemize}
    \item $\kleene{\Sigma}\times\kleene{\Sigma}\xrightarrow{m_{\Sigma}}\kleene{\Sigma}$
    \item $\terminal\times\kleene{\Sigma}\cong\kleene{\Sigma}$
    \item $\kleene{\Sigma}\times\terminal\cong\kleene{\Sigma}$
    \item because in a topos, for any object $A$, $A\times\initial\cong\initial\times A\cong\initial$, then if the source is $\initial$
      or the target is $\terminal$, then the composition is trivial
    \end{itemize}
\end{description}
\end{rmk}

\begin{prop}\label{prop:equiv_enriched_func_auto}
  Automata over $\Sigma$ are in bijective correspondence with $\eE$-functors $\uaA:\iword_\Sigma\xrightarrow{}\eE$ sending $(\objin,\objout)$ to $(\terminal,\classif)$
\end{prop}

\begin{proof}
  It is all about using the fact that $\iword_\Sigma$ is a free $\eE$-category. Given an automaton $(Q,i,F,\delta)$ over $\Sigma$, the $\eE$-quiver morphism $\alpha:Q_\Sigma\xrightarrow{}\eE$ is defined by
  \begin{itemize}
  \item $\alpha$ takes $(\objin,\objstates,\objout)$ to $(\terminal,Q,\classif)$
  \item $\alpha_{\objin,\objstates}=i:\terminal\xrightarrow{}\eE(\terminal,Q)\cong Q$
  \item $\alpha_{\objstates,\objout}=\nom{F}:\terminal\xrightarrow{}\eE(Q,\classif)=\classif^Q$ the global element corresponding to the subobject
    $F<Q$
  \item $\alpha_{\objstates,\objstates}=\delta^\vdash:\Sigma\xrightarrow{}\eE(Q,Q)=Q^Q$ the adjunct of
    $\delta:Q\times \Sigma\xrightarrow{}Q$, namely its corresponding morphism under the adjunction $(Q\times -)\dashv Q^{(-)}$
  \end{itemize}
  We obtain the wanted $\eE$-functor $\uaA:\iword_\Sigma\xrightarrow{}\eE$ as the adjunct of the $\eE$-quiver morphism $\alpha:Q_\Sigma\xrightarrow{}\eE$.
  Indeed, $\aA_0=\alpha_0$ so $\uaA$ sends $\objin$ to $\terminal$ and $\objout$ to $\classif$.

  Now given an $\eE$-functor $\uaA:\iword_\Sigma\xrightarrow{}\eE$ sending $\objin$ to $\terminal$ and $\objout$ to $\classif$, its adjunct
  $\eE$-quiver morphism $\alpha:Q_\Sigma\xrightarrow{}\eE$ is such that $\alpha(\objin)=\terminal$ and $\alpha(\objout)=\classif$
  and therefore we can define an automaton
  $$(\alpha(\objstates),\epsilon_\Sigma\alpha_{\objin,\objstates},F,\alpha_{\objstates,\objstates}^\dashv)$$ on $\Sigma$ in $\eE$
  where $F$ is the subobject of $\alpha_0(\objstates)$ corresponding to the characteristic morphism $\alpha_{\objstates,\objout}$.
\end{proof}

\begin{defi}
  The full sub-$\eE$-category of $\iword_\Sigma$ spanned by the objects $\objin$ and $\objout$ is denoted $\oword_\Sigma\xhookrightarrow{\iota_\Sigma}\iword_\Sigma$ and called the \emph{$\eE$-category of observable behaviors over the alphabet $\Sigma$}.
\end{defi}

\begin{prop}
  Under the bijection of Proposition \ref{prop:equiv_enriched_func_auto}, the restriction of a automaton $\uaA:\iword_\Sigma\xrightarrow{}\eE$ to the sub-$\eE$-category $\oword_\Sigma$ corresponds to the language $L(\aA)$ recognized by the complete deterministic automaton $\aA$ corresponding to $\uaA$.
\end{prop}

\begin{rmk}
  Observe first that the data of such an $\eE$-functor is entirely contained in its action on the object of
  morphisms between
  $\objin$ and $\objout$, namely $\underline{L}_{\objin,\objout}:\kleene{\Sigma}\xrightarrow{}\classif^\terminal\cong\classif$,
  which in turn is equivalent to the datum
  of a subobject of $\kleene{\Sigma}$, namely, a language on $\Sigma$.
\end{rmk}

\begin{proof}
  Consider an automaton $\aA=(Q,i,F,\delta)$ over $\Sigma$ as an $\eE$-functor $\uaA:\iword_\Sigma\xrightarrow{}\eE$. The language recognized by the former is  $$\terminal\xrightarrow{i}Q\xrightarrow{\delta^\ast}Q^{\kleene{\Sigma}}\xrightarrow{(\khi_F)^{\kleene{\Sigma}}}\classif^{\kleene{\Sigma}}$$ while the language recognized by the latter is the morphism $$\kleene{\Sigma}\xrightarrow{\aA_{\objin,\objout}}\classif$$ given by the restriction of $\uaA$ to the full sub-$\eE$-category $\oword_\Sigma$.
  
  Now $(\kleene{\Sigma},m_\Sigma,\epsilon_\Sigma)$ is a monoid so in particular, by left unit law,
  $\epsilon_\Sigma\times\kleene{\Sigma}:\kleene{\Sigma}\xrightarrow{}\kleene{\Sigma}\times\kleene{\Sigma}$
  is a section of the multiplication $m_\Sigma:\kleene{\Sigma}\times\kleene{\Sigma}\xrightarrow{}\kleene{\Sigma}$, so that the left triangle commute:
  \begin{center}  
    \begin{tikzcd}[ampersand replacement=\&]
      {\kleene{\Sigma}} \& {\kleene{\Sigma}} \& [+40pt] \classif \\ [+10pt]
      \& {\kleene{\Sigma}\times\kleene{\Sigma}} \& {Q^\times\classif^Q}
      \arrow["{\id_{\kleene{\Sigma}}}"', from=1-2, to=1-1]
      \arrow["{m_{\Sigma}}"', from=2-2, to=1-2]
      \arrow["{\comp_{\terminal,Q,\classif}}=\ev^Q_\classif"', from=2-3, to=1-3]
      \arrow["{\uaA_{\objin,\objstates}}", from=1-2, to=1-3]
      \arrow["{\uaA_{\objin,\objstates}\times\uaA_{\objstates,\objout}}", from=2-2, to=2-3]
      \arrow["{\epsilon\times\kleene{\Sigma}}"', from=1-1, to=2-2]
    \end{tikzcd}
  \end{center}
  and the right rectangle commute by the very definition of $\eE$-functoriality of $\uaA$.
  Therefore $\uaA_{\objin,\objout}$ is
  $$\kleene{\Sigma}\xrightarrow{\epsilon\times\id_{\kleene{\Sigma}}}\kleene{\Sigma}\times \kleene{\Sigma}
  \xrightarrow{\uaA_{\objin,\objstates}\times\uaA_{\objstates,\objout}}Q^\terminal\times\classif^Q
  \xrightarrow{\ev^Q_\classif}\classif$$
  and also
  $$\kleene{\Sigma}\xrightarrow{\epsilon\uaA_{\objin,\objstates}\times\uaA_{\objstates,\objout}}
  Q\times\classif^Q\xrightarrow{\ev^Q_\classif}\classif$$
  but $\epsilon\uaA_{\objin,\objstates}=i$ by definition of $\uaA$ as the adjunct of an $\eE$-quiver morphism.
  For the same reason, $\uaA_{\objstates,\objout}=\delta^\ast(\khi_F)^Q$, and finally
  $$\kleene{\Sigma}\xrightarrow{i\times(\delta^\ast(\khi_F)^Q)}Q\times\classif^Q\xrightarrow{\ev^Q_\classif}\classif$$
  is the adjunct of
  $$\terminal\xrightarrow{i}Q\xrightarrow{\delta^\ast}Q^{\kleene{\Sigma}}\xrightarrow{(\khi_F)^{\kleene{\Sigma}}}\classif^{\kleene{\Sigma}}$$
  because by naturality of $\ev^Q_\cdot:Q\times(-)^Q\xRightarrow{}\id_\eE$,
  \begin{center}
    \begin{tikzcd}[ampersand replacement=\&, outer sep=5pt]
      {Q\times Q^Q} \& [+30pt] {Q\times\classif^Q} \\
      Q \& \classif
      \arrow["{\ev^Q_Q}", from=1-1, to=2-1]
      \arrow["{\ev^Q_\classif}", from=1-2, to=2-2]
      \arrow["{\khi_F}", from=2-1, to=2-2]
      \arrow["{Q\times(\khi_F)^Q}", from=1-1, to=1-2]
    \end{tikzcd}
  \end{center}
  and because the adjunct $(\delta^\ast)^\vdash:Q\times\kleene{\Sigma}\xrightarrow{}Q$ is by definition
  $$Q\times\kleene{\Sigma}\xrightarrow{Q\times\delta^\ast}Q\times Q^Q\xrightarrow{\ev^Q_Q}Q$$
  so that the former morphism is
  $$\terminal\times\kleene{\Sigma}\xrightarrow{i\times\kleene{\Sigma}}
  Q\times\kleene{\Sigma}\xrightarrow{(\delta^\ast)^\vdash}Q\xrightarrow{\khi_F}\classif$$
  but finally, because of the naturality square
  \begin{center}
    \begin{tikzcd}[ampersand replacement=\&]
      {\eE(\kleene{\Sigma},Q^Q)} \& [+35pt] {\eE(\kleene{\Sigma},\classif^\terminal)} \\ [+15pt]
      {\eE(Q\times\kleene{\Sigma},Q)} \& {\eE(\kleene{\Sigma},Q^Q)}
      \arrow["{(-)^\vdash}", from=1-1, to=2-1]
      \arrow["{(-)^\vdash}", from=1-2, to=2-2]
      \arrow["{i;-;(\khi_F)^{\kleene{\Sigma}}}", from=1-1, to=1-2]
      \arrow["{(i\times\kleene{\Sigma});-;\khi_F}", from=2-1, to=2-2]
    \end{tikzcd}
  \end{center}
  where the vertical arrows are the adjunction isomorphisms, then this morphism has adjunct
  $$\terminal\xrightarrow{i}Q\xrightarrow{\delta^\ast}Q^{\kleene{\Sigma}}\xrightarrow{(\khi_F)^{\kleene{\Sigma}}}\classif^{\kleene{\Sigma}}$$
  which is (the global element corresponding to) the language recognized by $\aA$.
\end{proof}

\subsection{Category of automata}

There are several definitions of automata morphisms. We follow Colcombet and Petri\c{s}an \cite{colcombetAutomataMinimizationFunctorial2020}, i.e.\ the morphisms we consider are basically labeled quivers morphisms preserving the initial state and respecting final states.
This choice forces the languages recognized by two automata related by
such a morphism to be the same. It is not a problem here because the Myhill-Nerode theorem applies to a fixed, given language.

\begin{defi}
  Let $\underline{L}:\oword_\Sigma\xrightarrow{}\eE$ be a language over $\Sigma$ in a topos $\eE$.
  The category $\autocat(L)$ of \emph{L-automata} has as objects the $\eE$-functors extending $\underline{L}$ along the inclusion $\iota_\Sigma:\oword_\Sigma\xhookrightarrow{}\iword_\Sigma$, and as morphisms the morphisms of extensions of $\underline{L}$, namely $\eE$-natural transformations $\alpha:\uaA\xRightarrow{}\ubB$ between extensions of $\underline{L}$ restricting to the identity on $\underline{L}$.
\end{defi}

In some cases we will obtain an automaton recognizing the language only up to an automorphism of $\classif$, so that $\restr{\uaA}{\oword_\Sigma}$ is only isomorphic to $\underline{L}$. For example in $\sets$, not asking for strict equality would lead to having in $\autocat(L)$
both automata recognizing $L$ and automata recognizing the complement of $L$. But this is not an issue because we can always strictify:

\begin{lemma}[Strictification of an automaton with respect to a language]\label{lem:strictification-auto-wrt-lang}
  Let $\underline{L}$ be a language and $\uaA$ an automaton, both defined over $\Sigma$.
  If there exists an $\eE$-natural isomorphism $\phi:\restr{\uaA}{\oword_\Sigma}\cong \underline{L}$,
  then there exists an automaton $\ubB \in \autocat(L)$ isomorphic
  as an $\eE$-functor to $\uaA$ via $\psi:\uaA\cong\ubB$ such that $\iota_\Sigma\ast\psi=\phi$.
\end{lemma}

\begin{proof}
  Let $\ubB$ be the $\eE$-functor $\uaA$ with this only difference:
  $$\ubB_{\objstates,\objout} = \kleene{\Sigma}\xrightarrow{\aA_{\objstates,\objout}}\classif^Q\xrightarrow{(\phi_{\objout}^{-1})^Q}\classif^Q$$
  Equivalently, this is the $\eE$-functor defined on the free $\eE$-category $\iword_\Sigma$ by the $\eE$-quiver morphism $\beta$ where
  \begin{itemize}
  \item $\beta_0=\aA_0$
  \item $\beta_{\objin,\objstates}:\terminal\xrightarrow{\epsilon}\kleene{\Sigma}\xrightarrow{\uaA_{\objin,\objstates}}\uaA(\objstates)^{\uaA(\objin)}$
  \item $\beta_{\objstates,\objstates}:\Sigma\xhookrightarrow{}\kleene{\Sigma}\xrightarrow{\uaA_{\objstates,\objstates}}\uaA(\objstates)^{\uaA(\objstates)}$
  \item $\beta_{\objstates,\objout}:\terminal\xrightarrow{\epsilon}\kleene{\Sigma}\xrightarrow{\uaA_{\objstates,\objout}}\uaA(\objout)^{\uaA(\objstates)}
    \xrightarrow{(\phi_{\objout}^{-1})^{\uaA(\objstates)}}\uaA(\objout)^{\uaA(\objstates)}$
  \end{itemize}
\end{proof}

It was one of the main insights of Colcombet and Petri\c{s}an \cite{colcombetAutomataMinimizationFunctorial2020} that the minimal automaton recognizing a given language can be constructed by factoring the canonical map from the initial automaton to the final automaton. This remains true in our enriched context as we will see in the next section.

\subsection{Initial and terminal automata as enriched Kan extensions}

There is a whole theory of Kan extensions in the enriched case. See Kelly \cite{kellyBasicConceptsEnriched1982} and Loregian \cite{loregianCoendCalculus2021}
for excellent references on the subject.
In the unenriched case we have the following definition: consider a span of functors $\Cc'\xleftarrow{I}\Cc\xrightarrow{F}\Dd$, then a functor $R:\Cc'\xrightarrow{}\Dd$
is the right Kan extension $\Ran_I F$ of $F$ along $I$ if there exists an isomorphism
$$[\Cc,\Dd](IE,F)\cong[\Cc',\Dd](E,R)$$
natural in $E$.
Spelt out, it means that there exists a natural transformation $\rho:IR\xRightarrow{}F$ such that for
every other functor $E$ and natural transformation $\alpha:IE\xRightarrow{}F$, there exists a unique natural
transformation $\beta:E\xRightarrow{}R$ such that $\alpha=(I\ast\beta)\rho$. There also exists a dual notion of
left Kan extension.
This concept can be extended to an enriched setting and assuming the existence of certain weighted limits, we get pointwise formulas for right Kan extension resembling those occurring in the unenriched setting.

\begin{defi}
  Let $\vV=(\vV_0,\otimes, I)$ be a closed symmetric monoidal category, and let $\Cc' \xleftarrow{I} \Cc \xrightarrow{F} \Dd$
  be a diagram of $\vV$-functors.
  When the following weighted co/limits exist,
  we say the $\vV$-functor $\Ran_I(F)(-) := \lim^{\Cc'(-,I(=))}F$ is the \emph{pointwise right Kan extension} of $F$ along $I$,
  and that the $\vV$-functor $\Lan_I(F)(-) := \colim^{\Cc'(I(=),-)}F$ is the \emph{pointwise left Kan extension} of $F$ along $I$.
\end{defi}

We recall, in the appendix, how we can effectively compute enriched Kan extensions using conical co/limits and co/powers. In particular, Lemma \ref{lem:enriched-kan-ext-along-fully-faithful} ensures enriched Kan extension are (up to enriched natural isomorphism) enriched functors extensions when taken along a fully faithful enriched functor. This is important here because automata are extensions of the language they recognise along a full subcategory inclusion.

\begin{prop}\label{prop:init-term-autos-bicomp-topos}
  In a Grothendieck topos $\eE$, the initial and the terminal automata exist for any language $L$ over any alphabet $\Sigma$; they are respectively the left and the right $\eE$-enriched Kan extension of $\underline{L}:\oword_\Sigma\xrightarrow{}\eE$ along the fully faithful $\eE$-functor $\iota_\Sigma:\oword_\Sigma\xhookrightarrow{}\iword_\Sigma$.
\end{prop}

\begin{proof}
  Because of Lemma \ref{lem:end-existence}, enriched Kan extensions of $\underline{L}$ along $\iota_\Sigma$ exist,
  we denote $\emptyset(L)$ the left one and $\terminal(L)$ the right one.
  By Lemma \ref{lem:enriched-kan-ext-along-fully-faithful},
  the companion $\eE$-natural transformations of the $\eE$-Kan extension are isomorphisms.
  Finally, by Lemma \ref{lem:strictification-auto-wrt-lang},
  we can rename $\emptyset(L)$ and $\terminal(L)$ to be in the category $\autocat(L)$.
  Now those are respectively initial and terminal objects in the category $\autocat(L)$.
  Because of Lemma \ref{lem:strictification-auto-wrt-lang}, the unit of the left Kan extension $\initial(L)$
  is an automorphism $\lambda$ of $\underline{L}$.
  Therefore for all $\uaA\in\autocat(L)$, by Lemma \ref{lem:pt-kan-are-loc-kan} there exists a unique $\eE$-natural transformation
  $\alpha:\initial(L)\xRightarrow{}\uaA$ such that $\lambda(\iota_\Sigma\ast\alpha)=\lambda$,
  so $\iota_\Sigma\ast\alpha=\id_{\underline{L}}$, therefore there exists a unique
  automata morphism from $\initial(L)$ to $\uaA$. A dual reasoning apply for $\terminal(L)$.
\end{proof}

\begin{prop}\label{prop:explicit-init-term-autos}
  The initial and terminal automata can be computed explicitly:
  \begin{center}
    \begin{tabular}[]{l|r}
      $\initial(L)(\objstates)=\kleene{\Sigma}$ & $\terminal(L)(\objstates)=\classif^{\kleene{\Sigma}}$ \\
      $\initial(L)_{\objin,\objstates}=\id_{\kleene{\Sigma}}$ & $\terminal(L)_{\objin,\objstates}=(m_\Sigma\khi_L)^\dashv:\kleene{\Sigma}\xrightarrow{}\classif^{\kleene{\Sigma}}$ \\
      $\initial(L)_{\objstates,\objstates}={m_\Sigma}^\dashv:\kleene{\Sigma}\xrightarrow{} {\kleene{\Sigma}}^{\kleene{\Sigma}}$ & $\terminal(L)_{\objstates,\objstates}=((\classif^{m_{\kleene{\Sigma}}})^\vdash)^\vdash:\kleene{\Sigma}\xrightarrow{}(\classif^{\kleene{\Sigma}})^{(\classif^{\kleene{\Sigma}})}$\\
      $\initial(L)_{\objstates,\objout}:\kleene{\Sigma}\xrightarrow{}\classif^{\kleene{\Sigma}}$ & $\terminal(L)_{\objstates,\objout}=(\khi_{\ni_{\kleene{\Sigma}}})^\dashv:\kleene{\Sigma}\xrightarrow{}\classif^{\classif^{\kleene{\Sigma}}}$
    \end{tabular}
  \end{center}  
\end{prop}

  \begin{proof}
    Let us compute the terminal automaton. Its states object is defined as the end
    $$\int_{o\in\oword_\Sigma}\iword_\Sigma(\objstates,o)\pitchfork \underline{L}(o)$$
    so by Lemma \ref{lem:end-existence} it is obtained by equalizing
  \begin{center}
    \begin{tikzcd}[ampersand replacement=\&]
      {\prod_{o\in(\oword_\Sigma)_0}\underline{L}(o)^{\iword_\Sigma(\objstates,o)}} \& [+30pt] {\prod_{(o,o')\in(\oword_\Sigma)_0^2}\underline{L}(o')^{\iword_\Sigma(\objstates,o)\times\iword_\Sigma(o,o')}}
      \arrow["{\phi}", shift left=1.5, from=1-1, to=1-2]
      \arrow["{\psi}"', shift right=1.5, from=1-1, to=1-2]
    \end{tikzcd}
  \end{center}
  for the $\phi$ and $\psi$ of the lemma, which amounts to the equalizing of
  \begin{center}
    \begin{tikzcd}[ampersand replacement=\&]
      {\classif^{\kleene{\Sigma}}} \& [+30pt] {\classif^{\kleene{\Sigma}}}
      \arrow["{\id_{\classif^{\kleene{\Sigma}}}}", shift left=1.5, from=1-1, to=1-2]
      \arrow["{\id_{\classif^{\kleene{\Sigma}}}}"', shift right=1.5, from=1-1, to=1-2]
    \end{tikzcd}
  \end{center}
  because in each product, each other factor is $\terminal$ because either the exponent is $\initial$,
  either $\underline{L}(o)=\terminal$. Finally $\phi$ and $\psi$ happens to be both the identity morphism.

  \end{proof}

Now observe two things. First, the states objects of those automata witness the pointwise $\eE$-Kan extension expression
of the automata: for the initial one, $\kleene{\Sigma}$ is actually $\terminal=L(\objin)$ to the copower (in the enriched meaning)
$\kleene{\Sigma}=\iword_\Sigma(\objin,\objstates)$, and for the terminal one, $\classif^{\kleene{\Sigma}}$ is $\classif=L(\objout)$ to the power
$\kleene{\Sigma}=\iword_\Sigma(\objstates,\objout)$.
Second, those automata are never finite in the case $\eE=\sets$ if $\Sigma$ is non-empty. Their usefulness, with this respect, will be explained
in the next subsections: they let us compute the minimal automaton.

\subsection{Minimal automaton}

We understand minimal with respect to a given factorization system, following closely Colcombet and Petri\c{s}an \cite[Subsection 2.2]{colcombetAutomataMinimizationFunctorial2020}:
\begin{defi}
  In a category $\Cc$ endowed with an factorization system $(E,M)$,
  we say an object $X$ \emph{$(E,M)$-divides} an object $Y$
  if there exists a span $$X\xleftarrow{e\in E}Z\xrightarrow{m\in M}Y$$ in $\Cc$. An object is \emph{minimal} if it divides
  any object of $\Cc$.
\end{defi}

The case where divisibility is really meaningful is for $\Cc$ a regular category equipped with the $(\text{regular epimorphism},\text{monomorphism})$ factorization system, because then, morally, $X$ divides $Y$ if and only if $X$ is a
subquotient of $Y$. It makes sense in particular to consider divisibility for monoids. Any topos is a regular category in which every epimorphism is regular.

We recall the key idea of Colcombet and Petri\c{s}an \cite[Lemma 2.3]{colcombetAutomataMinimizationFunctorial2020} to compute the minimal automaton:

\begin{prop}\label{prop:ofs-init-term-min}
  Let $\Cc$ be a category with a factorization system $(E,M)$.
  If $\Cc$ has an initial and a terminal object, then the object through which the unique arrow from the initial to the terminal object $(E,M)$-factorizes is $(E,M)$-minimal.
\end{prop}

However, we need the factorization system on the category of automata which is a category of enriched functors. Therefore, we have to lift the $(\text{epi},\text{mono})$ factorization system on $\eE$ to $[\iword_\Sigma,\eE]_\eE$. Given two $\vV$-functors and a $\vV$-natural transformation $\alpha:\fF\xRightarrow{}\gG$, the pointwise factorization of $\alpha$ according to a given factorization system, might only give a unenriched functor. Thus, we need the factorization to have more properties, which leads to the definition of an enriched factorization system.

\begin{defi}
  Let $\vV$ be a symmetric closed monoidal category, and $\Cc$ a $\vV$-category.
  A factorization system $(E,M)$ on $\Cc$ is \emph{$\vV$-enriched} if for all $A\xleftarrow{e\in E}B$ and $X\xrightarrow{m\in M}Y$, the following square
  \begin{center}
    \begin{tikzcd}
      {\Cc(B,X)} & [+15pt] {\Cc(B,Y)} \\ [+15pt]
      {\Cc(A,X)} & {\Cc(A,Y)}
      \arrow["{\Cc(e,X)}"', from=1-1, to=2-1]
      \arrow["{\Cc(B,m)}", from=1-1, to=1-2]
      \arrow["{\Cc(e,Y)}", from=1-2, to=2-2]
      \arrow["{\Cc(A,m)}", from=2-1, to=2-2]
      \arrow["\lrcorner"{anchor=center, pos=0.125}, draw=none, from=1-1, to=2-2]
    \end{tikzcd}
  \end{center}
  is a pullback in $\vV$.
\end{defi}

One can characterize enriched factorization system amongst unenriched ones using powers or copowers, according to Lucyshyn-Wright \cite[Theorem 5.7]{lucyshyn-wrightEnrichedFactorizationSystems2014};
\begin{prop}
  If $\Cc$ has $\vV$-copowers (respectively $\vV$-powers), then a factorization system $(E,M)$ on $\Cc$ is enriched if and only if $E$ is stable under $\vV$-copowers (resp. $M$ is stable under $\vV$-powers).
\end{prop}

This is in particular the case if $\eE$ is a Grothendieck topos, and $(E,M)$ is the epi-mono factorization system.

In an unenriched context, a factorization system is in particular a functorial factorization system:
it defines a (strict) section of
the composition functor. An enriched factorization system has the same propriety: it defines a section of the composition enriched functor.
But all we need here is the fact that it allows for lifting a $\vV$-factorization system $(E,M)$ on a $\vV$-category $\Cc$ to any category of
enriched functors with target $\Cc$: the $\vV$-natural transformation are factorized pointwise with respect to $(E,M)$
and the objects through which they factorize are $\vV$-functorial.

\begin{prop}\label{prop:lift-enriched-ofs-on-enriched-functors}
  Let $(E,M)$ be an $\vV$-factorization system on a $\vV$-category $\Dd$, and $\Cc$ another $\vV$-category.
  Then consider the classes \[E_\Cc=\ens{\alpha:F\xRightarrow{}G\in\mor([\Cc,\Dd]_0)}{\forall X\in\Cc_0,\alpha_X\in E}\]
  of $\vV$-natural transformations that are pointwise in $E$,
  \[M_\Cc=\ens{\alpha:F\xRightarrow{}G\in\mor([\Cc,\Dd]_0)}{\forall X\in\Cc_0,\alpha_X\in M}\] of $\vV$-natural transformations
  that are pointwise in $M$. Then $(E_\Cc,M_\Cc)$ is a (unenriched) factorization system on $[\Cc,\Dd]$.
\end{prop}

\begin{proof}
  Consider two $\vV$-functors $F$ and $G$ from $\Cc$ to $\Dd$ and $\alpha$ a $\vV$-natural transformation from $F$ to $G$ which means that we have a collection $(\alpha_c:F(c)\xrightarrow{}G(c))_c$ of arrows of $\Dd_0$ (recall this means that $\alpha_c:I\xrightarrow{}\Dd(F(c),G(c))$ is an arrow in $\vV_0$) such that the diagram

  \begin{center}
    \begin{tikzcd}
      {\Cc(c,d)} & [+30pt] {\Dd(G(c),G(d))} \\ [+20pt]
      {\Dd(F(c),F(d))} & {\Dd(F(c),G(d))}
      \arrow["{G_{c,d}}"', from=1-1, to=2-1]
      \arrow["{F_{c,d}}", from=1-1, to=1-2]
      \arrow["{\Dd(F(c),\alpha_d)}", from=1-2, to=2-2]
      \arrow["{\Dd(\alpha_c,G(d))}", from=2-1, to=2-2]
    \end{tikzcd}
  \end{center}
  commute for all couple $(c,d)$ of objects of $\Cc$.
  Consider for all object $c$ of $\Cc$ the $(E,M)$-factorization $F(c)\xrightarrow{\epsilon_c}J_c\xrightarrow{\mu_c}G(c)$ of $\alpha_c$. The fact that the factorization system $(E,M)$ is enriched allows us to make $c\xmapsto{}J_c$ a $\vV$-functor:
  \begin{center}
    \begin{tikzcd}
      {\Cc(c,d)} & [+20pt] {\Dd(G(c),G(d))} & [+20pt] {} \\ [+20pt]
      {\Dd(F(c),F(d))} & {\Dd(J_c,J_d)} & {\Dd(J_c,G(d))} \\ [+20pt]
      {} & {\Dd(F(c),J_d)} & {\Dd(F(c),G(d))}
      \arrow["{F_{c,d}}", from=1-1, to=1-2]
      \arrow["{G_{c,d}}"', from=1-1, to=2-1]
      \arrow["{\Dd(F(c),\epsilon_d)}", from=1-2, to=2-3]
      \arrow["{\Dd(\mu_c,G(d))}"', from=2-1, to=3-2]
      \arrow["{J_{c,d}}", dotted, from=1-1, to=2-2]
      \arrow["{\Dd(\epsilon_c,J_d)}"', from=2-2, to=3-2]
      \arrow["{\Dd(J_c,\mu_d)}", from=2-2, to=2-3]
      \arrow["{\Dd(\epsilon_c,G(d))}", from=2-3, to=3-3]
      \arrow["{\Dd(F(c),\mu_d)}", from=3-2, to=3-3]
      \arrow["\lrcorner"{anchor=center, pos=0.125}, draw=none, from=2-2, to=3-3]
    \end{tikzcd}
  \end{center}
  where the outer hexagon commute because of $\vV$-naturality of $\alpha$, with $\alpha_c=\epsilon_c\mu_c$ and $\alpha_d=\epsilon_d\mu_d$. The $\vV$-functoriality of $J$ is due to the $\vV$-functoriality of $F$ and $G$, and the property of the pullback. We can recognize on this same diagram the $\vV$-naturality squares for $\mu$ and $\epsilon$, finally ensuring each arrow in $[\Cc,\Dd]$ admits an $(E_\Cc,M_\Cc)$-factorization.

  To conclude, $(E_\Cc,M_\Cc)$ is indeed a factorization system. The factorization is unique because it is unique pointwise, and a $\vV$-natural transformation is an isomorphism if and only if it is an isomorphism pointwise, so that $E_\Cc$ and $M_\Cc$ are closed under isomorphisms, and they are closed under composition too because so are $E$ and $M$, and because composition of $\vV$-natural transformations is also done pointwise.
\end{proof}

\begin{prop}\label{prop:ofs-lift-autocat}
  Any $\eE$-factorization system on $\eE$ can be lifted to $\autocat(L)$, so that the factorization of an automata morphism is obtained as the pointwise factorization of the underlying $\eE$-natural transformation.
\end{prop}

\begin{proof}
  Using Proposition \ref{prop:lift-enriched-ofs-on-enriched-functors}, we can lift an $\eE$-factorization system $(E,M)$ on $\eE$ to the $\eE$-category $[\iword_\Sigma,\eE]_0$.
  In particular, it gives a factorization system on this $\eE$-category seen as a category, and because $\autocat(L)$ is a subcategory
  of $[\iword_\Sigma,\eE]_0$, we have to show it induces a factorization system on $\autocat(L)$. It does because then $E_{\iword_\Sigma}\cap\autocat(L)$
  and $M_{\iword_\Sigma}\cap\autocat(L)$ are wide, replete subcategories of $\autocat(L)$. Given an automata morphism
  $\alpha:\uaA\xrightarrow{}\ubB$ and its factorization $\uaA\xRightarrow{\epsilon}\cC\xRightarrow{\mu}\ubB$ in $[\iword_\Sigma,\eE]_0$,
  then by construction of $E_{\iword_\Sigma}$ and $M_{\iword_\Sigma}$, we have
  \[\classif=\uaA(\objout)\xrightarrow{\epsilon_{\objout}}\cC(\objout)\xrightarrow{\mu_{\objout}}\classif=\ubB(\objout)=\id_\classif=\alpha_\classif\]
  same thing for $\objin$, but then by unicity of the factorization in $\eE_0$, $\cC(\objout)\cong\classif$ and $\cC(\objin)=\terminal$.
\end{proof}

\begin{defi}
  Let $L$ be a language on an alphabet $\Sigma$ in a bicomplete topos $\eE$.
  The automaton through which the unique arrow from the initial automaton to the terminal automaton factors is called the \emph{minimal automaton of $L$}, denoted $\Min(L)$.
\end{defi}

\begin{coro}\label{cor:minimal-automaton-existence}
  Let $L$ be a language on an alphabet $\Sigma$ in a bicomplete topos $\eE$.
  The minimal automaton of $L$ is minimal in $\autocat(L)$ with respect to $(\text{epi},\text{mono})$-divisibility, that is to say, $\Min(L)$ is a subquotient of any automaton that recognizes $L$. 
\end{coro}

\begin{proof}
  We can lift the $(\text{epi},\text{mono})$ factorization system of $\eE$ to $\autocat(L)$ according to Proposition \ref{prop:ofs-lift-autocat}, and this category has initial and terminal objects according to Proposition \ref{prop:init-term-autos-bicomp-topos}, so finally by Proposition \ref{prop:ofs-init-term-min} one obtain an automaton recognizing $L$ that is a subquotient of any automaton recognizing $L$.
\end{proof}

\subsection{Internal Nerode congruence}
\label{sec:intern-myhill-nerode}

In $\sets$, the \emph{Nerode congruence of $L$} is an equivalence relation associated with a language $L$ on an alphabet $\Sigma$ on words on $\Sigma$ defined by $u\sim_L v$ iff for all word x, $ux\in L \Leftrightarrow vx\in L$. It is strictly the same as saying $u\sim_L v$ iff $u^{-1}L = v^{-1}L$. Then the Nerode congruence is merely the kernel pair of left division of $L$, $u\xmapsto{}u^{-1}L$, which in turn is the adjunct of $(u,w)\xmapsto{} \khi_L(uw)$, namely the composite of monoid multiplication of $\kleene{\Sigma}$ and $\khi_L$.

\begin{prop}\label{prop:mor-from-0-to-1}
  In a bicomplete topos $\eE$, let $\uaA$ be an $L$-automaton over an alphabet $\Sigma$, the unique automata morphism $\alpha$ from $\initial(L)$ to $\uaA$ is given by
  $$\alpha_{\objstates} = \uaA_{\objin,\objstates} : \kleene{\Sigma} \xrightarrow{} Q$$
  and the unique automata morphism $\beta$ from $\uaA$ to $\terminal(L)$ by
  $$\beta_{\objstates} = (\kleene{\Sigma}\times Q \cong Q \times \kleene{\Sigma} \xrightarrow{\uaA_{\objstates,\objout}^\vdash} \classif)^\dashv : Q \xrightarrow{} \classif^{\kleene{\Sigma}}\text{.}$$
  In particular, the unique morphism from $\initial(L)$ to $\terminal(L)$ is $$(m_\Sigma\khi_L)^\dashv:\kleene{\Sigma}\xrightarrow{}\classif^{(\kleene{\Sigma})}\text{.}$$
\end{prop}

\begin{proof}
We already know that for any $\uaA\in\autocat(L)$, there is a unique automata morphism from $\initial(L)$
  to $\uaA$, we check
  $$(\uaA(\objin)\xrightarrow{\id_{\terminal}}\initial(L)(\objin),
  \uaA(\objstates)\xrightarrow{\uaA_{\objin,\objstates}}\initial(L)(\objstates),
  \uaA(\objout)\xrightarrow{\id_{\classif}}\initial(L)(\objout))$$
  is an $\eE$-natural transformation from $\initial(L)$ to $\uaA$. For example for the $\vV$-naturality square
  for the object of morphisms $\iword_\Sigma(\objstates,\objstates)$, consider the commuting diagram of respect
  of composition for the $\eE$-functor $\uaA$:
  \begin{center}
    \begin{tikzcd}[ampersand replacement=\&]
      {\iword_\Sigma(\objin,\objstates)\times\iword_\Sigma(\objstates,\objstates)} \& [+30pt] {\uaA(\objstates)^{\uaA(\objin)}\times\uaA(\objstates)^{\uaA(\objstates)}} \\ [+20pt]
      {\iword_\Sigma(\objin,\objstates)} \& {\uaA(\objstates)^{\uaA(\objstates)}}
      \arrow["{\comp_{\objin,\objstates,\objstates}}", from=1-1, to=2-1]
      \arrow["{\uaA_{\objin,\objstates}}", from=2-1, to=2-2]
      \arrow["{\uaA_{\objin,\objstates}\times\uaA_{\objstates,\objstates}}", from=1-1, to=1-2]
      \arrow["{\comp_{\uaA(\objin),\uaA(\objstates),\uaA(\objstates)}}", from=1-2, to=2-2]
    \end{tikzcd}
  \end{center}
  which is in fact
  \begin{center}
    \begin{tikzcd}[ampersand replacement=\&]
      {\kleene{\Sigma}\times\kleene{\Sigma}} \& [+30pt] {\uaA(\objstates)\times\uaA(\objstates)^{\uaA(\objstates)}} \\ [+20pt]
      {\kleene{\Sigma}} \& {\uaA(\objstates)}
      \arrow["{\uaA_{\objin,\objstates}}", from=2-1, to=2-2]
      \arrow["{m_{\Sigma}}", from=1-1, to=2-1]
      \arrow["{\ev_{\uaA(\objstates)}^{\uaA(\objstates)}}", from=1-2, to=2-2]
      \arrow["{\uaA_{\objin,\objstates}\times\uaA_{\objstates,\objstates}}", from=1-1, to=1-2]
    \end{tikzcd}
  \end{center}
  which in turn by the adjunction $(\kleene{\Sigma}\times -)\dashv (-)^{\kleene{\Sigma}}$ is
  \begin{center}
    \begin{tikzcd}[ampersand replacement=\&]
      {\kleene{\Sigma}} \& [+20pt] {\uaA(\objstates)^{\uaA(\objstates)}} \\ [+20pt]
      {{\kleene{\Sigma}}^{\kleene{\Sigma}}} \& {\uaA(\objstates)^{\kleene{\Sigma}}}
      \arrow["{(m_{\Sigma})^\dashv}", from=1-1, to=2-1]
      \arrow["{\uaA_{\objin,\objstates}}", from=1-1, to=1-2]
      \arrow["{\uaA(\objstates)^{\uaA_{\objin,\objstates}}}", from=1-2, to=2-2]
      \arrow["{(\uaA_{\objin,\objstates})^{\kleene{\Sigma}}}", from=2-1, to=2-2]
    \end{tikzcd}
  \end{center}
  which is the wanted $\eE$-naturality square.  
\end{proof}

\begin{defi}
  In a topos $\eE$ with a natural number object, let $\Sigma$ be an alphabet, and $L$ a language on $\Sigma$.
  The \emph{Nerode
    congruence\footnote{It is not an internal monoid congruence, it is only a categorical congruence, namely an internal
      equivalence relation.}
    of $L$}
  is the kernel pair of
  $$(m_\Sigma\khi_L)^\dashv:\kleene{\Sigma}\xrightarrow{}\classif^{(\kleene{\Sigma})}$$
  where $m_\Sigma$ is the internal concatenation of the free monoid $\kleene{\Sigma}$;
  in the internal language, the Nerode congruence is therefore
  $$\equiv_L=\ens{(x,y)\in (\kleene{\Sigma})^2}{\forall z \in \kleene{\Sigma}, x\cdot z\in L \Leftrightarrow y\cdot z\in L}$$
\end{defi}

\begin{prop}\label{prop:nerode-cong-min-auto}
  In a bicomplete topos $\eE$, the states object of $\Min(L)$ is the quotient of $\kleene{\Sigma}$ by the internal Nerode congruence
\end{prop}

\begin{proof}
  The quotient of $\kleene{\Sigma}$ by the Nerode congruence of $L$ is the coequalizer of the kernel pair of $(m_\Sigma\khi_L)^\dashv$ which in a regular category is canonically isomorphic to its image; but its image, in the $(\text{epi},\text{mono})$ factorization system, is the object through which $(m_\Sigma\khi_L)^\dashv$ factorizes, so that, by Corollary \ref{cor:minimal-automaton-existence} and Proposition \ref{prop:explicit-init-term-autos}, it is the states object of the minimal automaton.
\end{proof}

\subsection{Myhill-Nerode theorems for different finiteness conditions}

The following Myhill-Nerode theorems have two main cases of application: the first is in $\sets$, the classical Myhill-Nerode
theorem stating that a language is regular if and only if the Nerode congruence is of finite index,
and the second in the topos $\Nom$ of \emph{nominal sets}, proven for any $G$-sets topos by Boja\'nczyk, Klin and Lasota \cite{bojanczykAutomataTheoryNominal2014} (where $G$ is a discrete group or $G$ is the topological group of permutations of natural
numbers acting on discrete spaces), which states that a $G$-language (resp. nominal language) is regular, in the sense that it is recognized by an orbit-finite deterministic $G$-automaton (resp. nominal automaton) if and only if the quotient of the nominal set of words on the alphabet by the Nerode congruence is orbit-finite. Our Theorem \ref{thm:myhill-nerode-for-some-toposes} is a generalization and another point of view on Boja\'nczyk, Klin and Lasota \cite[Theorem 3.8]{bojanczykAutomataTheoryNominal2014}.

\begin{defi}
  Each time we consider a finiteness condition $(\textrm{FC})$, we say an automaton is $(\textrm{FC})$ if its states object is $(\textrm{FC})$.

  A language $L$ is $(\textrm{FC})$-regular if it admits an $(\textrm{FC})$ automaton that recognizes it.
\end{defi}

\begin{thm}\label{thm:myhill-nerode-for-some-toposes}
  Let $\eE$ be a bicomplete topos and $L$ a language on an alphabet $\Sigma$ of $\eE$.
  \begin{enumerate}
  \item For any non-empty class of points $P$ of $\eE$, $L$ is $P$-stalkwise-regular iff $\kleene{\Sigma}_{/\equiv_L}$ is $P$-stalkwise finite.
  \item If $\eE$ is an atomic Grothendieck topos, $L$ is decomposition-regular iff $\kleene{\Sigma}_{/\equiv_L}$ is decomposition-finite.
  \item If $\eE$ is Boolean, $L$ is K-regular iff $\kleene{\Sigma}_{/\equiv_L}$ is K-finite.
  \end{enumerate}
\end{thm}

\begin{proof}
  In the three cases, according to Proposition \ref{prop:stlk-fin-stable-subquo} and Proposition \ref{prop:fin-conds-stable-subquo},
  the finiteness conditions are preserved by subquotient. Because $\eE$ is bicomplete,
  the minimal automaton exists by Corollary \ref{cor:minimal-automaton-existence}.
  Now by Proposition \ref{prop:nerode-cong-min-auto}, $\kleene{\Sigma}_{/\equiv_L}$
  is the states object of $\Min(L)$ which divides any automaton recognizing $L$.
  So if some $\aA \in \autocat(L)$ has the finiteness
  condition, then so does $\Min(L)$, and therefore so does $\kleene{\Sigma}_{/\equiv_L}$.
\end{proof}

\subsection{The syntactic monoid}
\label{sec:syntactic-monoid}

There exists an algebraic notion of recognition where the recognizer is a monoid morphism. With this point of view, automata are merely presentations of such algebraic recognizers, given by the transition monoid of the automaton. Amongst monoids recognizing a language there is a smallest recognizer with respect to monoid divisibility: the syntactic monoid of a language. It can be defined abstractly as the quotient of the monoid of words by a ``syntactic'' congruence, or simply by the fact it is the transition monoid of the minimal automaton. We will now describe this (non-functorial) construction in any bicomplete topos with a natural number object and discuss its behavior with respect to a given finiteness condition.

\begin{defi}
  Let $L$ be a language on an alphabet $\Sigma$ in a topos $\eE$ with a natural number object.

  We say a monoid morphism $\phi:\kleene{\Sigma}\xrightarrow{}M$ \emph{recognizes $L$} if there exist $\khi:M\xrightarrow{}\classif$ making the following triangle commute:
  \begin{center}
    \begin{tikzcd}
      \kleene{\Sigma} & \classif \\
      M &
      \arrow["\khi_L", from=1-1, to=1-2]
      \arrow["\phi"', from=1-1, to=2-1]
      \arrow["\khi"', from=2-1, to=1-2]
    \end{tikzcd}
  \end{center}
  A monoid $M$ \emph{recognizes $L$} if there exists such a monoid morphism with target $M$.

  We call the triple $(M,\phi,\khi)$ an \emph{$L$-monoid} and an $L$-monoid morphism from $(M,\phi,\khi)$ to $(M',\phi',\khi')$ is a monoid morphism $f:M\xrightarrow{} M'$ such that those two triangles commute:
  \begin{center}
    \begin{tikzcd}
      & \kleene{\Sigma} & \\
      M & & M'
      \arrow["\phi"', from=1-2, to=2-1]
      \arrow["\phi'", from=1-2, to=2-3]
      \arrow["f", from=2-1, to=2-3]
    \end{tikzcd}
    and
    \begin{tikzcd}
      M & & M' \\
      & \classif &
      \arrow["f", from=1-1, to=1-3]
      \arrow["\khi"', from=1-1, to=2-2]
      \arrow["\khi'", from=1-3, to=2-2]
    \end{tikzcd}
  \end{center}
  in other words: $f$ is a morphism $\phi\xrightarrow{}\phi'$ in $\kleene{\Sigma}/\Mon(\eE)$ and $\khi\xrightarrow{}\khi'$ in $\eE/\classif$ as well. 
\end{defi}

\begin{rmk}
  If $\khi$ classifies $p:P\xhookrightarrow{}M$ and $\khi'$ classifies $q:Q\xhookrightarrow{}M'$, then we have $f\khi'=\khi$ iff $p=f^\ast(q)$.
\end{rmk}

\begin{lemma}\label{lem:refl-subcat-epic-lmon}
  If $\phi:\kleene{\Sigma}\xrightarrow{}M$ recognizes $L$, then $(\Im \phi,\phi:\kleene{\Sigma}\xrightarrow{}\Im \phi,\khi_{\phi(L)})$ is an $L$-monoid, where $\khi_{\phi(L)}$ is the characteristic morphism of the image inclusion of $L\xhookrightarrow{}\kleene{\Sigma}\xrightarrowdbl{\phi}\Im \phi$ in $\Im \phi$.

  This defines the reflector $\Im$ of the reflective full subcategory $\Sigma\Mon(L)$ of $\Sigma$-generated $L$-monoids, that is those $L$-monoids $(M,\phi,\khi)$ such that $\phi$ is epimorphic in $\eE$ and $\khi=\khi_{\phi(L)}$.
  This category can be equivalently described as the full subcategory of $\kleene{\Sigma}_{/\Mon(\eE)}$ spanned by $\phi:\kleene{\Sigma}\xrightarrowdbl{}M$ that are epimorphisms in $\eE$ and recognize $L$.
\end{lemma}

\begin{proof}
  By pasting law of pullbacks, because the outer rectangle is a pullback (the composite $\phi\khi$ is by hypothesis the characteristic morphism of $L$) and the right one two,
  \begin{center}
    \begin{tikzcd}
      L & & & \\
      & P & & \terminal \\
      \kleene{\Sigma} & & & \\
      & M & & \classif
      \arrow["!",from=1-1, to=2-4,bend left=20]
      \arrow["\exists !",from=1-1,to=2-2]
      \arrow["",from=1-1,to=3-1, hook]
      \arrow["!",from=2-2,to=2-4]
      \arrow["",from=2-2,to=4-2,hook]
      \arrow["\top",from=2-4,to=4-4]
      \arrow["\phi",from=3-1,to=4-2]
      \arrow["\khi",from=4-2,to=4-4]
      \arrow["\lrcorner"{anchor=center, pos=0.125}, draw=none, from=2-2, to=4-4]
    \end{tikzcd}
  \end{center}
  then by epi-mono factorization of $\phi$ and pulling back along the inclusion of $P$ we have a unique filler
  \begin{center}
    \begin{tikzcd}
      L & \phi(L) & P \\
      \kleene{\Sigma} & \Im\phi &  M
      \arrow["",from=1-1,to=1-2,two heads]
      \arrow["",from=1-1,to=2-1,hook]
      \arrow["",from=1-2,to=1-3,hook]
      \arrow["\exists !",from=1-2,to=2-2,hook]
      \arrow["",from=1-3,to=2-3,hook]
      \arrow["",from=2-1,to=2-2,two heads]
      \arrow["",from=2-2,to=2-3,hook]
    \end{tikzcd}
  \end{center}
  which is also a monomorphism as a pullback of a monomorphism. Then this diagram provides the epi-mono factorization of $L\xhookrightarrow{}\kleene{\Sigma}\xrightarrow{\phi}M$, and because all the squares here are pullbacks we have
  \begin{center}
    \begin{tikzcd}
      L & \phi(L) & \terminal \\
      \kleene{\Sigma} & \Im\phi &  \classif
      \arrow["",from=1-1,to=1-2,two heads]
      \arrow["",from=1-1,to=2-1,hook]
      \arrow["!",from=1-2,to=1-3]
      \arrow["",from=1-2,to=2-2,hook]
      \arrow["\top",from=1-3,to=2-3]
      \arrow["",from=2-1,to=2-2,two heads]
      \arrow["\khi_{\phi(L)}",from=2-2,to=2-3]
    \end{tikzcd}
  \end{center}
  showing the commutativity of the desired triangle by universal property of $\classif$.

  Now consider a $L$-monoid morphism $f:(M,\phi,\khi)\xrightarrow{}(N,\psi,\khi')$ with $\phi$ and $\psi$ epimorphic in $\eE$. We have to show that $f\khi_{\psi(L)}=\khi_{\phi(L)}$. For this consider the following diagram
  \begin{center}
    \begin{tikzcd}[ampersand replacement=\&]
      \& L \& L \& \\
      L \& \phi(L) \& \psi(L) \& \terminal \\
      \kleene{\Sigma} \& M \& N \& \classif
      \arrow[from=1-2,to=1-3,equal]
      \arrow[""{name=epiphil1},from=1-2,to=2-2,two heads]
      \arrow[""{name=epipsil},from=1-3,to=2-3,two heads]
      \arrow[from=epiphil1,to=epipsil,phantom,"\circlearrowleft"]
      \arrow[""{name=L},from=2-1,to=1-2,equal]
      \arrow[from=L,to=2-2,phantom,"\circlearrowleft",shift={(2pt,-2pt)}]
      \arrow[""{name=epiphil2},from=2-1,to=2-2, two heads]
      \arrow[""{name=1},from=2-1,to=3-1,hook]
      \arrow[from=2-2,to=2-3,two heads]
      \arrow[""{name=2},from=2-2,to=3-2,hook]
      \arrow["!",from=2-3,to=2-4]
      \arrow[""{name=3},from=2-3,to=3-3,hook]
      \arrow[""{name=4},"\top",from=2-4,to=3-4,hook]
      \arrow["\phi"',from=3-1,to=3-2,two heads]
      \arrow["f"',from=3-2,to=3-3,two heads]
      \arrow["\khi_{\psi(L)}"',from=3-3,to=3-4]
      \arrow[from=1,to=2, phantom, "\text{(1)}"]
      \arrow[from=2,to=3, phantom, "\text{(2)}"]
      \arrow[from=3,to=4, phantom, "\text{(3)}"]
    \end{tikzcd}
  \end{center}
  where the vertical composite morphisms $L\xrightarrow{}M$ and $L\xrightarrow{}N$ are the epi-mono factorization of, respectively $i\phi$ and $i\psi$ (where $i:L\xhookrightarrow{}\kleene{\Sigma}$ is the inclusion of $L$ in $\kleene{\Sigma}$), and $\phi(L)\xrightarrow{}\psi(L)$ is the unique filler. Showing $f\khi_{\psi(L)}=\khi_{\phi(L)}$ amount to showing the square (2+3) is a pullback. But we already know that (3) is a pullback so we have to show, by pullback pasting, that (2) is a pullback. But because $\eE$ is regular, $\phi$ is an epimorphism and (1) is a pullback (because $\phi\khi_{\phi(L)}=\khi_L$), then (2) is a pullback iff (1+2) is (according to Carboni, Janelidze, Kelly and Par\'e \cite[Lemma 4.6]{carboniLocalizationStabilizationFactorization1997}). But $\phi f=\psi$ so (1+2) is indeed a pullback, translating $\psi\khi_{\psi(L)}=\khi_L$.
\end{proof}

Of course, $\kleene{\Sigma}$ always recognizes any language $L$. This can be seen as the consequence of the fact that the initial automaton always exists (in the cocomplete case).

\begin{defi}
  Let $\uaA:\iword_\Sigma\xrightarrow{}\Set$ be an automaton on an alphabet $\Sigma$ in a topos $\eE$ with a natural number object.
  The morphism $\uaA_{\objstates,\objstates}:\kleene{\Sigma}\xrightarrow{}\uaA(\objstates)^{\uaA(\objstates)}$ is a monoid morphism with image factorization $\kleene{\Sigma}\xrightarrowdbl{\tau_{\uaA}}T(\aA)\xhookrightarrow{}\uaA(\objstates)^{\uaA(\objstates)}$. The monoid $T(\aA)$ is called the \emph{transition} monoid of $\aA$.   
\end{defi}

\begin{prop}
  If $\uaA$ recognizes $L$, then $(T(\uaA),\tau_{\uaA})$ is $\Sigma$-generated $L$-monoid.
\end{prop}

\begin{proof}
  We apply Lemma \ref{lem:refl-subcat-epic-lmon} to the following triangle
  \begin{center}
        \begin{tikzcd}
      & [+10pt] \kleene{\Sigma} & [+10pt] \\ [+10pt]
      \uaA(\objstates)^{\uaA(\objstates)} & & \classif
      \arrow["\uaA_{\objstates,\objstates}"', from=1-2, to=2-1]
      \arrow["\uaA_{\objin,\objout}=\khi_L", from=1-2, to=2-3]
      \arrow["((\epsilon_\Sigma\uaA_{\objstates,\objout})^\vdash)^{\epsilon_\Sigma\uaA_{\objin,\objstates}}"', from=2-1, to=2-3]
    \end{tikzcd}
  \end{center}
  which commutes because of $\eE$-functoriality of $\uaA$ and the fact that $\epsilon_\Sigma$ is the identity of $\objstates$ in the $\eE$-category $\iword_\Sigma$.
\end{proof}

\begin{rmk}
  However, the $T$ construction is not functorial; to witness this in $\Set$, consider any finite automaton $\aA$ with at least two distinct states $q$ and $r$ on an alphabet with at least two letters $a$ and $b$, and construct an automaton $\bB$ by adding a new state $t$ to $\aA$, and such that $\ubB(a)(t)=q$, $\ubB(b)(t)=r$, $\ubB(c)(t)=t$ if $c\in A\setminus\point{a,b}$ and $\ubB(c)(s)=\uaA(c)(s)$ if $c\in \Sigma, s\in\uaA(\objstates)$. The initial state and final states of $\bB$ are those of $\aA$ so that the inclusion of states of $\aA$ in those of $\bB$ defines a monomorphic automata morphism from $\aA$ to $\bB$, and the transition monoid of $\bB$ contains strictly more endofunctions than those of $\aA$. However, an $L$-monoid morphism between $\tau_\aA$ and $\tau_\bB$ has to be surjective because $\tau_\aA$ and $\tau_\bB$ are, which is impossible in that case.
\end{rmk}

The transition monoid construction might not be functorial but it at least preserves divisibility.

\begin{prop}\label{prop:trans-mon-preserv-div}
  Let $L$ be a language in a topos with a natural number object.
  If $\aA\in\autocat(L)$ divides $\bB\in\autocat(L)$, then $T(\aA)$ divides $T(\bB)$.
  Better, $T$ restricted to the wide subcategory $\autocat_{\text{epi}}(L)$ of $\autocat(L)$ of automata and pointwise epic automata morphisms is a covariant functor, and $T$ restricted to the wide subcategory $\autocat_{\text{mono}}(L)$ of $\autocat(L)$ of automata and pointwise monic automata morphisms is a contravariant functor.
\end{prop}

\begin{proof}
  Consider a pointwise epimorphic automata morphism $e:\uaA\xrightarrow{}\ubB$. By $\eE$-naturality of $e$, epimorphy of $e$ entailing monomorphy of ${\ubB(\objstates)}^e$ and epi-mono factorization we have a unique filler
    \begin{center}
    \begin{tikzcd}
      \kleene{\Sigma} & T(\uaA) & & \uaA(\objstates)^{\uaA(\objstates)} \\
      \kleene{\Sigma} & T(\ubB) & \ubB(\objstates)^{\ubB(\objstates)} & \ubB(\objstates)^{\uaA(\objstates)}
      \arrow["\tau_{\uaA}",from=1-1,to=1-2,two heads]
      \arrow["",from=1-1,to=2-1,equal]
      \arrow["",from=1-2,to=1-4,hook]
      \arrow["\exists !T(e)",from=1-2,to=2-2,two heads]
      \arrow["e^{\uaA(\objstates)}",from=1-4,to=2-4]
      \arrow["\tau_{\ubB}",from=2-1,to=2-2,two heads]
      \arrow["",from=2-2,to=2-3,hook]
      \arrow["{\ubB(\objstates)}^e",from=2-3,to=2-4,hook]
    \end{tikzcd}
  \end{center}
  making the diagram commute and it also is an epimorphism. By functoriality of orthogonal factorization systems, this construction is functorial where it makes sense, namely on $\autocat_{\text{epi}}(L)$.

  Consider now a pointwise monomorphic $m:\uaA\xrightarrow{}\ubB$. By the same sort of arguments, notably because monomorphy of $m$ implies monomorphy of $m^{\uaA(\objstates)}$, we have a unique filler
  \begin{center}
    \begin{tikzcd}
      \kleene{\Sigma} & T(\ubB) & & \ubB(\objstates)^{\ubB(\objstates)} \\
      \kleene{\Sigma} & T(\uaA) & \uaA(\objstates)^{\uaA(\objstates)} & \ubB(\objstates)^{\uaA(\objstates)}
      \arrow["\tau_{\ubB}",from=1-1,to=1-2,two heads]
      \arrow["",from=1-1,to=2-1,equal]
      \arrow["",from=1-2,to=1-4,hook]
      \arrow["\exists !T(m)",from=1-2,to=2-2,two heads]
      \arrow["{\ubB(\objstates)}^m",from=1-4,to=2-4]
      \arrow["\tau_{\uaA}",from=2-1,to=2-2,two heads]
      \arrow["",from=2-2,to=2-3,hook]
      \arrow["m^{\uaA(\objstates)}",from=2-3,to=2-4,hook]
    \end{tikzcd}
  \end{center}
  and this construction is contravariantly functorial from $\autocat_{\text{mono}}(L)$ to $\Sigma\Mon(L)$.
\end{proof}

An automaton recognizing $L$ can be seen as a presentation of an $L$-monoid. But in fact, each $\Sigma$-generated $L$-monoid can be seen canonically as the transition monoid of an automaton.

\begin{lemma}\label{lem:sect-trans-mon-epi}
  The covariant functor $T:\autocat_{\text{epi}}(L)\xrightarrow{}\Sigma\Mon(L)$ has a section (up to natural isomorphism) $A$ defined by
  $$\underline{A(M,\phi)}(\objstates)=M\text{,}$$ $$\underline{A(M,\phi)}_{\objin,\objstates}=\phi\text{,}$$  $$\underline{A(M,\phi)}_{\objstates,\objstates}=\kleene{\Sigma}\xrightarrow{\phi}M\xrightarrow{m^\dashv}M^M\text{ and}$$ $$\underline{A(M,\phi)}_{\objstates,\objout}=\kleene{\Sigma}\xrightarrow{\phi}M\xrightarrow{m^\dashv}M^M\xrightarrow{\khi^M}\classif^M$$ where $m$ is the multiplication of the monoid $M$.
\end{lemma}

\begin{proof}
  First, note that this indeed defines a functor sending an $L$-monoid morphism $f:(M,\phi)\xrightarrow{}(N,\psi)$ to the automata morphism defined by $f$.
  
  Now we have to show that $M$ is isomorphic to a submonoid of $M^M$ in a natural way; it is sort of an internal Cayley theorem.
  
  Recall that by definition $\kleene{\Sigma}\xrightarrowdbl{\tau_{\underline{A(M,\phi)}}}T(\underline{A(M,\phi)})\xhookrightarrow{}M^M$ is the regular epi-mono factorization of the monoid morphism  $\underline{A(M,\phi)}_{\objstates,\objstates}=\phi m^\dashv$. But at the same time $\phi$ is epic and $M^e$, where $e$ is the unit of the monoid $M$, is a retract of $m^\dashv$ by left unitality. Therefore by unicity of the factorization, $T(\underline{A(M,\phi)})$ and $M$ are isomorphic as monoids under $\kleene{\Sigma}$.
\end{proof}

\begin{thm}
  Let $L$ be a language in a bicomplete topos $\eE$ with a natural number object.
  The transition monoid of the minimal automaton $T(\Min(L))$ is minimal in the category of $L$-monoids. We then call this monoid the \emph{syntactic monoid of $L$} and denote it $\Syn(L)$.
\end{thm}

\begin{proof}
  First, bicompleteness ensures the very existence of the minimal automaton according to Corollary \ref{cor:minimal-automaton-existence}. Consider any $L$-monoid $(M,\phi,\khi)$, by Lemma \ref{lem:refl-subcat-epic-lmon} it has an epimorphic sub-$L$-monoid $(\Im\phi,\phi)$ which can be sent to an automaton $A(\Im\phi)$ which is divided by the minimal automaton $\Min(L)$. By Proposition \ref{prop:trans-mon-preserv-div}, $\Syn(L):=T(\Min(L))$ then divides $T(A(\Im\phi))$ which in turn is (isomorphic to) $\Im\phi$ by Lemma \ref{lem:sect-trans-mon-epi}. Because $\Im\phi$ is a sub-$L$-monoid of $M$, it in particular divides it, so that finally, $\Syn(L)$ divides $M$.
\end{proof}

In $\Set$, the syntactic monoid provides another characterization of regularity: a language is regular if and only if its syntactic monoid is finite. We discuss this fact with different finiteness conditions.

In the following results we might use the fact that $\Min(L)$ is a \emph{reachable automaton}.

\begin{defi}
  An $L$-automaton $\uaA$ is \emph{reachable} if $\initial(L)\xrightarrow{!}\uaA$ is an epimorphism.
\end{defi}

\begin{lemma}\label{lem:st-obj-quo-trans-mon-reachable}
  If $\uaA$ is reachable then $\tau_{\uaA}=\uaA_{\objstates,\objstates}$ so that in particular the states object of $\uaA$ is a quotient of its transition monoid.
\end{lemma}

\begin{proof}
Let $Q$ be the states object of $\uaA$ and consider the following diagram
    \begin{center}
      \begin{tikzcd}
        \terminal\times\kleene{\Sigma} & \kleene{\Sigma}\times\kleene{\Sigma} & [+25pt] Q\times Q^Q \\
        & \kleene{\Sigma} & Q
        \arrow["\epsilon_\Sigma\times\kleene{\Sigma}",from=1-1,to=1-2]
        \arrow["\uaA_{\objin,\objstates}\uaA_{\objstates,\objstates}",from=1-2,to=1-3]
        \arrow["",hook,two heads,from=1-1,to=2-2]
        \arrow["m_\Sigma",from=1-2,to=2-2]
        \arrow["\uaA_{\objin,\objstates}",from=2-2,to=2-3]
        \arrow["\ev^Q_Q",from=1-3,to=2-3]
      \end{tikzcd}
    \end{center}
    where the left triangle commute by unitality in the monoid $\kleene{\Sigma}$ and the right square does because of $\eE$-functoriality of $\uaA$. The outer pentagon has adjoint
    \begin{center}
      \begin{tikzcd}
        \kleene{\Sigma} & Q^Q \\
        & Q^\terminal
        \arrow["\uaA_{\objstates,\objstates}",from=1-1,to=1-2]
        \arrow["\uaA_{\objin,\objstates}"',from=1-1,to=2-2,two heads]
        \arrow["Q^{\epsilon_\Sigma\uaA_{\objin,\objstates}}",from=1-2,to=2-2,two heads]
      \end{tikzcd}
    \end{center}
    where $Q^{\epsilon_\Sigma\uaA_{\objin,\objstates}}$ is epic because of the adjunction $Q^{(-)}\dashv Q^{(-)}:\eE\op\xrightarrow{}\eE$. Then $\uaA_{\objstates,\objstates}$ is an epimorphism so that $Q^Q=T(\uaA)$ and this show that there is an epimorphism from the transition monoid to the states object of $\uaA$.

\end{proof}

\begin{thm}
  Let $L$ be a language in a bicomplete topos $\eE$.
  If $\Syn(L)$ is K-finite then $L$ is K-regular.
  
  If $\eE$ is Boolean and $L$ is K-regular then $\Syn(L)$ is K-finite.
\end{thm}

\begin{proof}
  It suffices to show that $\Min(L)$ is K-finite if $\Syn(L)$ is K-finite.
  If $\Syn(L)$ is K-finite, then by definition $\Min(L)$ is reachable so by Lemma \ref{lem:st-obj-quo-trans-mon-reachable}, the states object of $\Min(L)$ is a quotient of a K-finite object, so it is itself K-finite.
  
  If $\eE$ is Boolean and $L$ is K-regular, then by Theorem \ref{thm:myhill-nerode-for-some-toposes} $\Min(L)$ is K-finite. Then $\Min(L)^{\Min(L)}$ has to be K-finite according to Acu\~na-Ortega and Linton \cite[Main Theorem]{acuna-ortegaFinitenessDecidability1979}. Now $\Syn(L)$ being a complemented subobject of $\Min(L)^{\Min(L)}$, is K-finite by Berger and Iwaniack \cite[Lemma 3.5]{bergerProfiniteFundamentalGroup2023}.
\end{proof}

\subsection{Geometric morphism lifting}
\label{sec:geom-morph-lift}

Inverse images of geometric morphisms should, in some sense, transport automata as they are models of first order logic theories. In fact, we can lift the underlying adjunction of a geometric morphism to the corresponding categories of automata, given a language in the target.

\begin{defi}
  Consider $F:\eE\xrightarrow{}\fF$ a left exact functor between elementary toposes (notably an inverse or direct image of a geometric morphism).
  We define the canonical morphisms:
  \begin{itemize}
  \item $F(\classif_\eE)\xrightarrow{}\classif_\fF$ classifying the subobject $\terminal\cong F(\terminal)\xhookrightarrow{F(\top)}F(\classif_\eE)$ (recall a left exact functor preserves by definition finite limits therefore monomorphisms),
  \item for objects $A$ and $B$ of $\fF$, $F(B^A)\xrightarrow{}F(B)^{F(A)}$ adjoint of $F(A)\times F(B^A)\cong F(A\times B^A)\xrightarrow{F(\ev^A_B)}F(B)$
  \end{itemize}

\end{defi}

\begin{lemma}\label{lem:inv-im-preserv-free-mon}
  Let $f:\eE\xrightarrow{}\fF$ be a geometric morphism between two toposes such that $\fF$ admits a natural number object. For every language $L$ in $\fF$ on an alphabet $\Sigma$, $f^\ast(\kleene{\Sigma})$ is the free monoid on $f^\ast(\Sigma)$ so that $f^\ast(L)$ is a language on $f^\ast(\Sigma)$.
\end{lemma}

\begin{proof}
  The fact that $f^\ast(\kleene{\Sigma})$ is a monoid is because $f^\ast$ preserves finite limits, and same thing for $f_\ast$, so that the adjunction $f^\ast\dashv f_\ast$ can be lifted to the categories of monoid objects in the two corresponding toposes:
    \begin{align*}
      \Mon(\eE)(f^\ast(\kleene{\Sigma}),M) &\cong \Mon(\fF)(\kleene{\Sigma},f_\ast(M)) \\
                                      &\cong \fF(\Sigma,f_\ast(M)) \\
                                      &\cong \eE(f^\ast(\Sigma),M)
    \end{align*}
  therefore $f^\ast(\kleene{\Sigma})$ is (isomorphic to) the free monoid generated by $f^\ast(\Sigma)$.
\end{proof}

\begin{lemma}\label{lem:inv-im-lift}
  The inverse image $f^\ast$ can be lifted to automata categories
  \begin{center}
    \begin{tikzcd}
      \autocat(L) & \autocat(f^\ast(L)) \\
      \fF & \eE
      \arrow["f^\ast_{\autocat}",from=1-1,to=1-2]
      \arrow["\objstates",from=1-1,to=2-1]
      \arrow["\objstates",from=1-2,to=2-2]
      \arrow["f^\ast",from=2-1,to=2-2]
    \end{tikzcd}
  \end{center}
by sending $\uaA:\iword_\Sigma\xrightarrow{}\fF$ to $f^\ast_{\autocat}(\uaA):\iword_{f^\ast(\Sigma)}\xrightarrow{}\eE$ defined by $$f^\ast_{\autocat}(\uaA) \text{ sends }(\objin,\objstates,\objout)\text{ to }(\terminal, f^\ast(\uaA(\objstates)),\classif_\eE)\text{,}$$
$$f^\ast_{\autocat}(\uaA)_{\objin,\objstates}:f^\ast(\kleene{\Sigma})\xrightarrow{f^\ast(\uaA_{\objin,\objstates})}f^\ast(\uaA(\objstates))\text{,}$$
$$f^\ast_{\autocat}(\uaA)_{\objstates,\objstates}:f^\ast(\kleene{\Sigma})\xrightarrow{f^\ast(\uaA_{\objstates,\objstates})}f^\ast(\uaA(\objstates)^{\uaA(\objstates)})\xrightarrow{}f^\ast(\uaA(\objstates))^{f^\ast(\uaA(\objstates))}\text{ and}$$
$$f^\ast_{\autocat}(\uaA)_{\objstates,\objout}:f^\ast(\kleene{\Sigma})\xrightarrow{f^\ast(\uaA_{\objstates,\objout})}f^\ast(\classif_\fF^{\uaA(\objstates)})\xrightarrow{}\classif_\eE^{\uaA(\objstates)}$$
and sending an automata morphism $\alpha:\uaA\xrightarrow{}\ubB$ to $f^\ast(\alpha)$.
\end{lemma}
  \begin{proof}
  We check that $f^\ast_{\autocat}(\uaA)$ is an automaton by checking it is an $\eE$-functor. For example, it respects endomorphisms composition at $\objstates$ as witnessed by the diagram:
  \begin{center}
    \begin{tikzcd}
      f^\ast(\kleene{\Sigma})^2 & & [+40pt] & [-20pt] &  [-20pt] (f^\ast(Q)^{f^\ast(Q)})^2 \\ 
      & & & f^\ast(Q^Q)\times f^\ast(Q^Q) & \\  [+20pt]
      & f^\ast(\kleene{\Sigma}\times\kleene{\Sigma}) & f^\ast(Q^Q\times Q^Q) & & \\ [+10pt]
      f^\ast(\Sigma) & & f^\ast(Q^Q) & & f^\ast(Q)^{f^\ast(Q)}

      \arrow["m_{f^\ast(\kleene{\Sigma})}",from=1-1,to=4-1]
      \arrow["\sim",from=1-1,to=3-2]
      \arrow["{f^\ast_{\autocat}(\uaA)_{\objstates,\objstates}\times f^\ast_{\autocat}(\uaA)_{\objstates,\objstates}}",from=1-1,to=1-5]

      \arrow["\comp",from=1-5,to=4-5]

      \arrow["c\times c",from=2-4,to=1-5]

      \arrow["f^\ast(m_{\kleene{\Sigma}})",from=3-2,to=4-1]
      \arrow["f^\ast(\uaA_{\objstates,\objstates}\times\uaA_{\objstates,\objstates})",from=3-2,to=3-3]

      \arrow["f^\ast(\comp)",from=3-3,to=4-3]
      \arrow["\sim",from=3-3,to=2-4]

      \arrow["f^\ast(\uaA_{\objstates,\objstates})",from=4-1,to=4-3]

      \arrow["c",from=4-3,to=4-5]

    \end{tikzcd}
  \end{center}
  commuting because the left triangle does by the first part of the proof, the bottom left square does because of $\eE$-naturality of $\uaA$, the top pentagram does by preservation of product by $f^\ast$, and finally the bottom right does because of ``coherency'' of the canonical morphisms.
\end{proof}

\begin{lemma}\label{lem:dir-im-lift}
  In a similar fashion, the direct image can be lifted to a functor $f_{\ast,\autocat}:\autocat f(^\ast(L))\xrightarrow{}\autocat(L)$ by setting for all $f^\ast(L)$-automaton $\ubB$ the $L$-automaton $f_{\ast,\autocat}(\ubB)$ defined by $$f_{\ast,\autocat}(\ubB) \text{ sends }(\objin,\objstates,\objout)\text{ to }(\terminal, f_\ast(\ubB(\objstates)),\classif_\fF)\text{,}$$
$$f_{\ast,\autocat}(\ubB)_{\objin,\objstates}=(\ubB_{\objin,\objstates})^\dashv: \kleene{\Sigma}\xrightarrow{}f_\ast(\ubB(\objstates))\text{,}$$
$$f_{\ast,\autocat}(\ubB)_{\objstates,\objstates}:\kleene{\Sigma}\xrightarrow{(\ubB_{\objin,\objstates})^\dashv}f_\ast(\ubB(\objstates)^{\ubB(\objstates)})\xrightarrow{}f_\ast(\ubB(\objstates))^{f_\ast(\ubB(\objstates))}\text{ and}$$
$$f_{\ast,\autocat}(\ubB)_{\objstates,\objout}:\kleene{\Sigma}\xrightarrow{(\ubB_{\objin,\objstates})^\dashv}f_\ast(\classif_\eE^{\ubB(\objstates)})\xrightarrow{}\classif_\fF^{\ubB(\objstates)}$$
  
\end{lemma}

\begin{thm}\label{thm:geom-lift-auto}
  Let $f:\eE\xrightarrow{}\fF$ be a geometric morphism between elementary toposes such that $\fF$ admits a natural number object, and let $L$ be a language over $\Sigma$ in $\fF$. The underlying adjunction $(f^\ast\dashv f_\ast):\eE\xrightarrow{}\fF$ can be lifted to the categories of automata $(f^\ast_{\autocat}\dashv f_{\ast,\autocat}):\autocat(L)\xrightarrow{}\autocat f^\ast(L)$ with morphism of adjunction the evaluation at $\objstates$ object, and the left adjoint preserves divisibility of automata.
\end{thm}

\begin{proof}
  The whole lifting is the consequence of Lemma \ref{lem:inv-im-preserv-free-mon}, Lemma \ref{lem:inv-im-lift} and Lemma \ref{lem:dir-im-lift}. As for the preservation of divisibility, it is due to the fact that $f^\ast$ does preserve it and it is the action on automata morphism of $f^\ast_{\autocat}$.
\end{proof}

\section{Examples}
\subsection{Equivariant automata}
\label{sec:equiv-autom}

\begin{prop}
  We have a functor $\espcl:\Grp\xrightarrow{}\Geom$ defined by $\espcl G=[G,\Set]$ and for all group homomorphism $f:G\xrightarrow{}H$, and the inverse image of the geometric morphism $\espcl f$ sends a $H$-set to its ``restriction'' along $f$, and $\espcl f$ is in fact essential.
  
  In particular, for $H=\terminal$, $\espcl f$ is the global section morphism, $\espcl f_\ast(A)$ is the set of fixed points of $A$ and $\espcl f_!(A)$ is its set of orbits, while if only $G=\terminal$, then $\espcl f^\ast$ is the forgetful functor, $\espcl f_!(S)=S\times H$ is the $S$-copower of $H$ as a $H$-set, and $\espcl f_\ast(S)=S^H$ where $S^H$ is endowed with the action $\alpha\cdot h(h')=\alpha(h'h^{-1})$.
  This last geometric morphism is by definition a point, and it is the unique (up to isomorphism) point of $[H,\Set]$.
\end{prop}

\begin{proof}
  In terms of $H$-sets seen as presheaves $F$, $\espcl f^\ast(F)=fF$ where $f$ is seen as a functor. Then, limits and colimits being computed pointwise in $[H,\Set]$, $\espcl f^\ast$ has to preserve both.
  
  One way of seeing there is only one point of $[H,\Set]$ is by Diaconescu theorem: points of $[H,\Set]$ are flat functors $H\op\xrightarrow{}\Set$, which means the corresponding left $H$-set has to be simply transitive, and up to isomorphism, there is only one simply transitive left $H$-set, namely $H$.
\end{proof}

\begin{defi}
  An equivariant automaton is an automaton in a topos $\espcl G$ for $G$ a discrete group.
\end{defi}

\begin{prop}\label{prop:fin-in-gset}
  Let $G$ be a discrete group. An object $A$ of $\espcl G$ is
  \begin{enumerate}
  \item dK-finite iff it is finite as a set
  \item decomposition-finite iff it has a finite number of orbits
  \end{enumerate}
\end{prop}

\begin{proof}
  (2) is immediate while (1) is a general fact where $G$ can be replaced by any small groupoid, see Johnstone \cite[Example 5.4.19]{johnstoneSketchesElephant2002}.
\end{proof}

\begin{coro}
  Let $f:G\xrightarrow{}H$ be any group homomorphism, and $L$ be a language over $\Sigma$ in $[H,\Set]$, we can lift $f$ to the categories of equivariant automata by lifting $\espcl f$ according to Theorem \ref{thm:geom-lift-auto}, so that if $L$ is dK-regular (respectively decomposition-regular), then so is $f(L)=(L,-\cdot f(=))$.
  
  In particular, for $f=!:\terminal\xrightarrow{}H$, if $L$ is dK-regular then it is regular in the classical meaning.
\end{coro}

\begin{example}
  As a first ``toy'' example we consider an automaton in the topos of sets with an involution, namely $\zZ_{/2\zZ}\Set$, the topos of the actions of the two-element group. For each set with an involution $(X,i)$ we shall denote $i(x)=\overline{x}$. 

  Consider the two-letter alphabet $\Sigma=\point{a,\overline{a}}$ where the involution exchanges the two letters. The free (internal) monoid is simply the free monoid $\kleene{\Sigma}$ where the involution swaps the two letters.
  We define the language $L=\ens{lu\overline{l}}{l\in A, u\in A^\ast}$ of words of length at least two whose first and last letters are different. The Nerode quotient $\kleene{\Sigma}_{/\equiv_L}$ is the five elements set $$\point{L,a^{-1}L,\overline{a}^{-1}L,(a\overline{a})^{-1}L,(\overline{a}a)^{-1}L}$$ so that its only fixed point is $L\in \kleene{\Sigma}_{/\equiv_L}$. This allows us to describe the minimal $\zZ/2\zZ\Set$-automaton of this $\zZ/2\zZ\Set$-language:

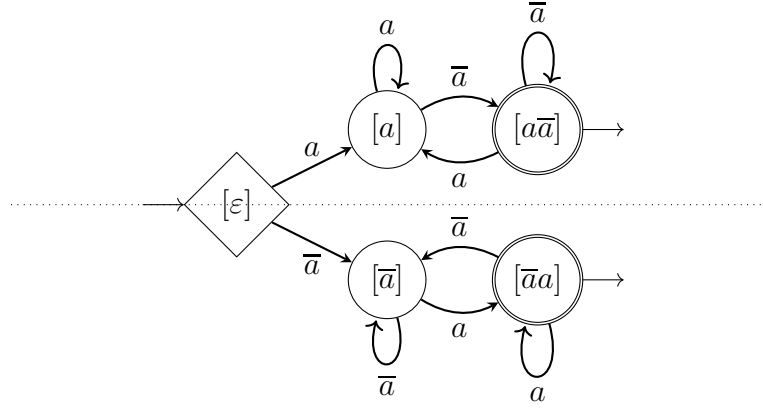
  \begin{center}
    \begin{tikzpicture} [node distance = 2cm, on grid]
      \draw[dotted] (-3,0) -- (7,0);
      \node (init) [state,initial, initial text = {}, initial by diamond] at (0,0) {$[\epsilon]$};
      \node (a) [state] at (2,1){$[a]$};
      \node (abar) [state] at (2,-1){$[\overline{a}]$};
      \node (aabar) [state,accepting, accepting text = {}, accepting right] at (4,1) {$[a\overline{a}]$};
      \node (abara) [state,accepting, accepting text = {}, accepting right] at (4,-1) {$[\overline{a}a]$};

      \path [-stealth, thick]
      (init) edge node [above] {$a$} (a)
      (a) edge [loop above] node {$a$} ()
      (a) edge [bend left] node [above] {$\overline{a}$} (aabar)
      (aabar) edge [bend left] node [below] {$a$} (a)
      (aabar) edge [loop above] node {$\overline{a}$} ()
      
      (init) edge node [below] {$\overline{a}$} (abar)
      (abar) edge [loop below] node {$\overline{a}$} ()
      (abar) edge [bend right] node [below] {$a$} (abara)
      (abara) edge [bend right] node [above] {$\overline{a}$} (abar)
      (abara) edge [loop below] node {$a$} ();
    \end{tikzpicture}
  \end{center}

  Observe the symmetry with respect to the dotted line which gives the involution on the set of states. We can go further and compute the syntactic monoid of the language, which is a monoid object in $\zZ_{/2\zZ}\Set$; it happens to have five elements, and is in fact the rectangular band of type $2\times 2$ to which we freely added a neutral element to make it a monoid:
  \begin{center}
    \begin{tabular}{|c|c|c|c|c|c|}
      \hline
      $\downarrow\cdot\rightarrow$ & $\epsilon$ & $a$ & $a\overline{a}$ & $\overline{a}$ & $\overline{a}a$ \\
      \hline
      $\epsilon$ & \cellcolor{gray!5}$\epsilon$ & \cellcolor{gray!5}$a$ & \cellcolor{gray!5}$a\overline{a}$ & \cellcolor{gray!5}$\overline{a}$ & \cellcolor{gray!5}$\overline{a}a$ \\
      \hline
      $a$ & \cellcolor{gray!5}$a$ & \cellcolor{gray!20}$a$ & \cellcolor{gray!20}$a\overline{a}$ & \cellcolor{gray!20}$a\overline{a}$ & \cellcolor{gray!20}$a$ \\
      \hline
      $a\overline{a}$ & \cellcolor{gray!5}$a\overline{a}$ & \cellcolor{gray!20}$a$ & \cellcolor{gray!20}$a\overline{a}$ & \cellcolor{gray!20}$a\overline{a}$ & \cellcolor{gray!20}$a$ \\
      \hline
      $\overline{a}$ & \cellcolor{gray!5}$\overline{a}$ & \cellcolor{gray!20}$\overline{a}a$ & \cellcolor{gray!20}$\overline{a}$ & \cellcolor{gray!20}$\overline{a}$ & \cellcolor{gray!20}$\overline{a}a$ \\
      \hline
      $\overline{a}a$ & \cellcolor{gray!5}$\overline{a}a$ & \cellcolor{gray!20}$\overline{a}a$ & \cellcolor{gray!20}$\overline{a}$ & \cellcolor{gray!20}$\overline{a}a$ & \cellcolor{gray!20}$\overline{a}a$ \\
      \hline
    \end{tabular}
  \end{center}

  Recall the rectangular band of type $2\times 2$ is the semigroup on the set $\point{0,1}\times\point{0,1}$ where the multiplication is defined by $(s,t)\cdot (x,y)=(s,y)$. It is in fact a $\zZ_{/2\zZ}\Set$-rectangular band where the involution swaps $0$ and $1$.
\end{example}
\subsection{Continuous automata}
\label{sec:etales-autom}


Let $B$ be any topological space.

\begin{defi}
  A continuous language is a language in a topos of sheaves over a topological space.
  
  A continuous automaton is an automaton in a topos of sheaves over a topological space.

  A continuous automaton is stalkwise finite if its state object is for $P$ the set of points of $B$, and a continuous language is stalk-regular if it admits a stalkwise finite automaton recognizing it.
\end{defi}

\begin{thm}[{Theorem \ref{thm:myhill-nerode-for-some-toposes} for stalkwise finiteness}]
  Let $L$ be a language over an alphabet $\Sigma$ in $\Sh(B)$. It is stalk-regular iff the quotient sheaf $\kleene{\Sigma}_{/\equiv_L}$ is stalkwise finite.
\end{thm}
  

\begin{defi}[{Berger and Iwaniack \cite[Definition 3.1]{bergerProfiniteFundamentalGroup2023}}]
  A sheaf is \emph{finite} if it is both dK-finite and decomposition-finite in $\Sh(B)$.
  We denote $\Sh(B)_{\text{sf}}$ the full subcategory spanned by coproducts of finite sheaves in $\Sh(B)$.
\end{defi}

\begin{thm}[{Berger and Iwaniack \cite[Theorem 3.11]{bergerProfiniteFundamentalGroup2023}}]
  $\Sh(B)_{\text{sf}}$ is an atomic Grothendieck topos.
\end{thm}

\begin{coro}
  If $\Sigma$ belongs to $\Sh(B)_{\text{sf}}$, then $L$ is regular iff $\kleene{\Sigma}_{/\equiv_L}$ is finite.
\end{coro}

\begin{proof}
  Because $\Sigma$ is in $\Sh(B)_{\text{sf}}$, everything happens in this topos where we have a both points of Theorem \ref{thm:myhill-nerode-for-some-toposes} that apply.
\end{proof}


The following theorem is the fundamental theorem of covering spaces:
\begin{thm*}
  If $B$ is semi-locally simply connected, then $LC(B)\cong\pi_1(B)\sets$ where $\pi_1(B)$ is the fundamental groupoid of $B$.
\end{thm*}

\begin{thm}
  Let $B$ be semi-locally simply connected, $\Sigma$ be a locally constant sheaf, and $L$ a language over $\Sigma$.
  \begin{enumerate}
  \item $L$ is decomposition-regular iff $\kleene{\Sigma}_{/\equiv_L}$ is decomposition-finite.
  \item $L$ is dK-regular iff $\kleene{\Sigma}_{/\equiv_L}$ is stalkwise finite for a point in each connected component of $B$.
  \end{enumerate}
\end{thm}

\begin{proof}
  This is just the application of both points of Theorem \ref{thm:myhill-nerode-for-some-toposes} because $\pi_1(B)\sets$ is an atomic Grothendieck topos and therefore Boolean. The fact that in point (2) the ``dK-finite'' condition is replaced by a ``stalkwise finite'' condition is due to proof of Proposition \ref{prop:fin-in-gset}, because dK-finiteness in a topos of groupoid actions $LC(B)\cong\pi_1(B)\sets$ boils down to finiteness of the underlying sets for each connected component of the groupoid (which, in this case, are the connected components of the space $B$).
\end{proof}


\begin{thm}
  Let $B$ be any topological space, $\Sigma$ be a coproduct of locally constant sheaves, and $L$ a language over $\Sigma$.
  \begin{enumerate}
  \item $L$ is decomposition-regular iff $\kleene{\Sigma}_{/\equiv_L}$ is decomposition-finite.
  \item $L$ is dK-regular iff $\kleene{\Sigma}_{/\equiv_L}$ is dK-finite.
  \end{enumerate}
\end{thm}

\begin{proof}
  According to Theorem \ref{thm:leroy}, $SLC(B)$ is an atomic Grothendieck topos and $\Sh(B)$ is connected over it, and because $\Sigma$ is in $SLC(B)$, then Theorem \ref{thm:myhill-nerode-for-some-toposes} is applied in this topos.
\end{proof}

\subsection{Nominal automata}
\label{sec:nom-autom}
We define the Myhill-Schanuel topos $\espcl\Aut(\noms)$ of \emph{nominal sets} and \emph{equivariant} functions to be the category of continuous actions on discrete spaces (i.e.\ sets) of the topological group $\Aut_\noms$ permutations of a countable set $\noms$ of \emph{names}, where the topology is induced by the inclusion $\Aut(\noms)\subset \noms^\noms$  (i.e.\ the product topology of infinitely-many discrete spaces, or the simple convergence topology); equivariant functions are those functions that commute with the action. According to Mac Lane and Moerdijk \cite[Theorem 3.9.2]{maclaneSheavesGeometryLogic1994}, this indeed form an atomic Grothendieck topos.

\begin{defi}
  A nominal automaton is an automaton in the Myhill-Schanuel topos $\espcl\Aut(\noms)$.
\end{defi}

We will give different characterizations of the Myhill-Schanuel topos. One of them is a direct translation of what it means to be continuous for a $\Aut(\noms)$-set, and makes use of the notion of the support of an element:

\begin{lemma}\label{lem:cont-action-iff-stab-neigh-id}
  Let $G$ be a topological group.
  \begin{enumerate}
  \item A subgroup $H$ of $G$ is open if and only if it is a neighborhood of the neutral element.
  \item An action $E\times G\xrightarrow{a} E$ of $G$ on a discrete space $E$ is continuous if and only if each isotropy group $\stab(x)=\ens{g\in G}{x\cdot g=x}$ for $x\in E$ is a neighborhood of $e$.
  \end{enumerate}
\end{lemma}

\begin{proof}
  Note that in $G$, left (respectively right) multiplication by $g\in G$ is open, because it has a continuous inverse, namely left (resp.\ right) multiplication by $g^{-1}$.
  
  First, if $H$ is a subgroup of $G$ that is a neighborhood of $e$, then for an open subset $\vV$ such that $e\in\vV\subset H$ we have $$H=\bigcup_{h\in H}h\vV$$ that is open because by the note $h\vV$ is open.

  Second, suppose each isotropy group is a neighborhood of $e$, it is open by the previous point, so that for all $x\in E$,
  $$a^{-1}(x)=\ens{(x\cdot g, h)}{(x\cdot g)\cdot h=x}=\bigcup_{g\in G}\point{x\cdot g}\times g^{-1}\stab(x)$$
  is open because $E$ is discrete so $\point{x\cdot g}$ is open and $g^{-1}\stab(x)$ is open as well.
  If $a$ is continuous, then for all $x\in E$, $a^{-1}(x)$ is open so that $\stab(x)=\pi_2(a^{-1}(x)\bigcap\point{x}\times G)$ is open too, where $\pi_2:E\times G\xrightarrow{} G$ is the open, canonical projection out of the product space.
\end{proof}

\begin{prop}\label{prop:permut-topol-basis}
  The topology of $\Aut(\noms)$ admits the sets $\noms_{F,j}$ of permutations whose restriction to $F$ is the injection $j:F\xhookrightarrow{}\noms$ as a basis, for any finite subset $F$ of $\noms$ and injection $j$.
\end{prop}

\begin{proof}
  A basis of $\prod_{a\in\noms}\noms$ is given by sets $\prod_{a\in \noms}S_a$ where $S_a=\noms$ for $a\in\noms$ but for a finite subset $F$ of $\noms$, where $S_a$ is an arbitrary subset of $\noms$, because the topology on $\noms$ is discrete, and for this same reason we can even restrict to the case where those $S_a$ are singletons, so the basis can be restricted to set $\ens{g:\noms\xrightarrow{}\noms}{\restr{g}{F}=f}$ where $F$ is any finite subset of $\noms$ and $f:F\xrightarrow{}\noms$ is any function. If we restrict it to $\Aut(\noms)\subset\prod_{a\in\noms}\noms$, then we see a base of $\Aut(\noms)$ is given by sets $\ens{\sigma\in\Aut(\noms)}{\restr{\sigma}{F}=f}$ for any finite subset $F$ of $\noms$ and $f:F\xrightarrow{}\noms$ is any function, but those sets are non-empty if and only if $f$ is injective.
\end{proof}

\begin{defi}
  Let $E$ be an $\Aut(\noms)$-set, and let $S\subset \noms$. We say $S$ \emph{supports} $x$ if any permutation of $\noms$ that fixes $S$ stabilizes $x$. Moreover if $S$ is finite, we say $S$ is a \emph{finite support} of $x$. 
\end{defi}

\begin{coro}\label{cor:aut-set-cont-iff-all-elem-fs}
  An $\Aut(\noms)$-set is continuous if and only each of its element admits a finite support.
\end{coro}

\begin{proof}
  According to Proposition \ref{prop:permut-topol-basis}, a neighborhood basis of $\id_\noms$ is given by sets
  $$\ens{\sigma\in\Aut(\noms)}{\forall a\in F, \sigma(a)=a}$$
  where $F$ is any finite subset of $\noms$. So by Lemma \ref{lem:cont-action-iff-stab-neigh-id}, an $\Aut(\noms)$-set is continuous if and only if the isotropy group of each element contains a set of the form $\ens{\sigma\in\Aut(\noms)}{\forall a\in F, \sigma(a)=a}$, which is equivalent to saying $F$ is a finite support of this element.
\end{proof}

\begin{prop}
  The topos $\espcl\Aut(\noms)$ is equivalent to
  \begin{enumerate}
  \item the full subcategory of the category of (non-necessarily continuous) $\Aut(\noms)$-sets spanned by $\Aut(\noms)$-sets $E$ whose elements are \emph{finitely supported}.
  \item the category of pullback preserving functors from the category of finite sets and injections to the category of sets
  \end{enumerate}
\end{prop}

\begin{proof}
  The first characterization is Corollary \ref{cor:aut-set-cont-iff-all-elem-fs}.

  By Mac Lane and Moerdijk \cite[Theorem 3.9.2]{maclaneSheavesGeometryLogic1994}, $\espcl\Aut(\nN)$ is a Grothendieck topos that can be presented by the atomic site of the full subcategory of transitive $\Aut(\nN)$-sets, or its dense atomic subsite given by $\Aut(\nN)$-sets of the form $\Aut(\ens{i\in\nN}{i\geq n})=\ens{\sigma\in\Aut(\nN)}{\forall i<n, \sigma(i)=i}$ for any $n\in\nN$ and equivariant functions, which in turn is equivalent to the atomic site of the dual of finite sets and injections. Then, the sheaf condition is exactly preservation of pullbacks in this case.
  
\end{proof}

\begin{prop}
  The category $\espcl\Aut(\noms)$ of nominal sets is coreflective inside $[\Aut(\noms),\Set]$, where the coreflector keeps only the finitely supported elements of a $\Aut(\noms)$-set. In particular, arbitrary colimits and finite limits in $\espcl\Aut(\noms)$ are computed pointwise while arbitrary limits are computed by applying the coreflection to the pointwise limit.
\end{prop}

\begin{proof}
  We can manually check that finite limits are computed as in $\Set$. Then, the forgetful functor $\mathcal{U}:\espcl\Aut(\noms)\xrightarrow{}\Set$ is the inverse image of a geometric morphism (more precisely, a point): it preserves finite limits and its right adjoint maps a set $S$ to the set of finitely supported functions $f:\Aut(\noms)\xrightarrow{}S$ with respect to the action $f\cdot \sigma = (\pi\xmapsto{}f(\pi\sigma))$. Because $\espcl\Aut(\noms)$ is a Grothendieck topos, it has all limits and colimits, and $\mathcal{U}$ preserves all the latter while only the finite former.
\end{proof}

\begin{prop}
  Let $E$ and $F$ be nominal sets.
  \begin{itemize}
  \item The object classifier is the two element set with trivial $\Aut(\noms)$ action.
  \item The exponential $F^E$ is the set of finitely supported functions $f:E\xrightarrow{}F$ with respect to the action $(f\cdot\sigma)(x)=f(x\cdot\sigma^{-1})\cdot \sigma$.
  \item In particular the power object $\classif^E$ is the set of finitely supported subsets of $E$ with respect to the action $P\cdot\sigma=\ens{x\cdot\sigma}{x\in P}$.
  \end{itemize}
\end{prop}

The set of ``names'' $\noms$ is itself, canonically, a nominal set. Moreover, it is decomposition-finite as it is transitive.

\begin{example}
Consider on this alphabet $\noms$ the classical example of the language $L$ of words where the first letter appears at least once again further in the word: $$L=\ens{ab_1b_2\cdots b_n \in \kleene{\noms}}{n\in\nN,a,b_i\in\noms,\exists i\in \nN, 1\leq i\leq n, b_i=a}$$ which is a nominal set. Indeed, it is stable under permutations of letters, and each word is finitely supported by the finite set of letters that appears in it.

Let us compute the minimal automaton for this language. Recall that for any nominal set $A$, $\classif^A=\ens{P\subset A}{P\text{ is finitely supported}}$. The states object is $\kleene{\noms}/\equiv_L$ and here it is therefore the nominal set $$\ens{u^{-1}L\subset\kleene{\noms}}{u\in\kleene{\noms}}=\point{L}\cup\ens{\kleene{\noms}a\kleene{\noms}}{a\in\noms}\cup\point{\kleene{\noms}}$$ where each set of this union is an orbit, so that by Theorem \ref{thm:myhill-nerode-for-some-toposes}, $L$ is decomposition-regular.

Now to finally describe the minimal automaton, recall that the initial state, a fixed point, is simply the equivalence class $L$ of $\epsilon$, and an equivalence class is a final state if and only if it contains a language that contains the empty string, $\epsilon$. The only such class is $\kleene{\noms}$, therefore it is the only final state of the automaton. Then, the action $-\cdot a$ of a letter $a\in\noms$ is given by $K\cdot a = a^{-1}K$. The following diagram sums up the construction
and the register automaton counterpart of this nominal automaton can be found in Francez and Kaminski \cite[Figure 7]{kaminskiFinitememoryAutomata1994}:

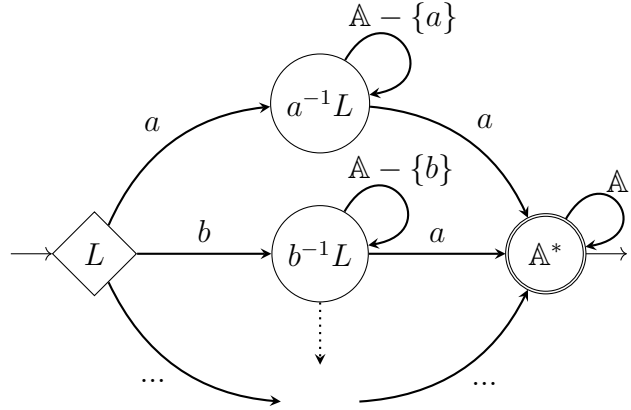
\begin{center}
  \begin{tikzpicture} [node distance = 2cm, on grid]
    \node (i) [state,initial, initial text = {}, initial by diamond] at (0,0) {$L$};
    \node (b) [state] at (3,0) {$b^{-1}L$};
    \node (a) [state] at (3,2){$a^{-1}L$};
    \node (z) [state, draw = none] at (3,-2) {};
    \node (f) [state,accepting, accepting text = {}, accepting right] at (6,0) {$\kleene{\noms}$};

    \path [-stealth, thick, every loop/.style={min distance=10mm,in=10,out=60,looseness=5}]
    (i) edge [bend left] node [above left] {$a$} (a)
    (i) edge node [above] {$b$} (b)
    (i) edge [bend right] node [below left] {$...$} (z)

    (a) edge [loop above] node {$\noms-\point{a}$} ()
    (b) edge [loop above] node {$\noms-\point{b}$} ()
    
    (a) edge [bend left] node [above right] {$a$} (f)
    (b) edge node [above] {$a$} (f)
    (z) edge [bend right] node [below right] {$...$} (f)

    (f) edge [loop above] node {$\noms$} ()
    
    (b) edge [dotted] node {} (z);
  \end{tikzpicture}
\end{center}

where the diamond state is the initial state and the double circle is a final state (in fact the only one in this case). Observe that states in the same column are in the same orbit. The orbit $\ens{a^{-1}L}{a\in\noms}$ can be thought of as a single state such that a transition from the initial state to this state-orbit writes the read letter (which is the first letter of the word) in a register. Reading the rest of the word, we loop on this state-orbit until we read a letter that is other than the one in the register. In that case we reach the final state on which we loop until the word is finished reading.

Let us compute the syntactic monoid of $L$. It is the set of functions $\ens{u^{-1}L}{u\in\kleene{\noms}}\xrightarrow{}\ens{u^{-1}L}{u\in\kleene{\noms}}$ of the form $f_v:u^{-1}L\xmapsto{} (uv)^{-1}L$ for some fixed $v$ and with action $f_v\cdot\sigma=f_{v\cdot\sigma}$ because $v\xmapsto{}f_v$ is a nominal monoid morphism $\kleene{\noms}\xrightarrowdbl{}\Syn(L)$ so in particular an equivariant function.
Then $$\Syn(L)=\point{f_{\epsilon}}\cup\ens{f_a}{a\in\noms}\cup\ens{f_{abu}}{a\in\noms,b\in\noms\setminus\point{a},u\in\kleene{\noms\setminus\point{a}}}\cup\ens{f_{auav}}{a\in\noms,u,v\in\kleene{\noms}}$$
and each member of this union is an orbit, so that $\Syn(L)$ is decomposition-finite.
\end{example}

\begin{prop}
  The dK-finite objects in $\espcl\Aut(\noms)$ are exactly finite sets with the trivial action.
\end{prop}

\begin{proof}
  The topos $\espcl\Aut(\noms)$ is a subtopos of $[\Aut(\noms),\Set]$, so dK-finite nominal sets have to be dK-finite $\Aut(\noms)$-sets, and according to Proposition \ref{prop:fin-in-gset}, those have a finite underlying set. But non-trivial transitive nominal sets have an infinite underlying set; according to \cite[Theorem 6.3]{bojanczykSlightlyInfiniteSets2019}, transitive nominal sets $A$ are of the form ${\noms^{[F]}}_{/G}$ where $F$ is a (least) finite support of an element of $A$, $\noms^{[F]}$ is the set of injections from $F$ to $\noms$, and $G$ a subgroup of $\permut_F$. But then the only case for which $\noms^{[F]}_{/\permut_F}$ is a finite set is $F\cong\terminal$, so that $A=\terminal$.
\end{proof}

In the topos $\espcl\Aut(\noms)$, point 2 of Theorem \ref{thm:myhill-nerode-for-some-toposes} becomes:
\begin{thm}[Boja\'nczyk {\cite[Theorem 3.8]{bojanczykAutomataTheoryNominal2014}}]
  Let $L$ be a language over $\Sigma$ in the topos $\espcl\Aut(\noms)$. The language $L$ is decomposition-regular (i.e.\ recognized by an orbit-finite complete deterministic nominal automaton) iff $\kleene{\Sigma}_{/\equiv_L}$ is orbit-finite. 
\end{thm}

\section{Conclusion and future perspectives}
\label{sec:conclusion}

Because the subobject classifier in $\Set$ played a crucial r\^ole in the Colcombet and Petri\c{s}an functorial viewpoint of automata , we adapted it to a wide class of toposes, recovering minimization results, and adding Myhill-Nerode type theorems to it, as well as some discussions around the syntactic monoid of a language, everything internally to a given topos. The results still make sense for sets, and can be applied to the Myhill-Schanuel topos of nominal sets.\newline

We would like to make more use of the enriched Colcombet and Petri\c{s}an functorial point of view of automata. Amongst the work to be done there is:
\begin{itemize}
\item Finding other examples of toposes where studying automata is meaningful (other than the already known $\sets$ and $\espcl\Aut(\noms)$).
\item Enriching in other monoidal categories, for example the categories of Ad\'amek, Milius and Urbat \cite{adamekSyntacticMonoidsCategory2015},
  which are monoidal closed as Kleisli categories for monads on $\sets$ (therefore canonically strong) or
  as Eilenberg-Moore categories for commutative monads. Specifically, $\Rel$ is the Eilenberg-Moore category for the powerset monad
  which is commutative, $\Par$ is the Kleisli category of the \emph{Option} monad $(-)+\terminal$, and the first category allow
  for speaking of non-deterministic automata and the second of deterministic, non necessarily complete automata.
\item Treating the case of Brzozowski's algorithm for nominal automata, which won't work in the sense that that the powerset construction
  doesn't preserve orbit-finiteness: indeed there is no chance the powerset of an orbit-finite $G$-set for $G$ an infinite group be
  orbit-finite... Also, it is known that non-deterministic nominal automata are strictly more expressive than the deterministic one,
  for there exist languages recognized by orbit-finite non-deterministic automata that are not recognized by any deterministic nominal automata.
\item Treating the case of Choffrut's algorithm for nominal transducers.
\end{itemize}

\appendix


\section{Pointwise enriched Kan extensions}
\label{sec:pw-enrich-kan-ext}

When $\Cc'$ has ``enough'' co/powers, there exists a description of Kan extensions using ends and coends:

$$ \Ran_I(F)(-) = \int_{x:\Cc'}\Cc'(-,I(x))\pitchfork F(x) $$
and
$$ \Lan_I(F)(-) = \int^{x:\Cc'}\Cc'(I(x),-)\odot F(x) $$
and in turn those ends can be computed using co/products, co/equalizers and co/powers as it will be shown in the next lemma.

We introduce some notation. When a $\vV$-category $\Dd$ has powers, then it means we have a $\vV$-natural isomorphism
$$\vV(v,\Dd(d,d')) \cong \Dd(d,v\pitchfork d')$$
in $v$, $d$ and $d'$, which restricts to a natural isomorphism
$$\vV_0(v,\Dd(d,d'))\cong \Dd_0(d,v\pitchfork d')$$
and we have the dual facts for copowers; by definition we have a $\vV$-natural isomorphism
$$\vV(v,\Dd(d,d')) \cong \Dd(v\odot d,d')$$
in $v$, $d$ and $d'$, which restricts to a natural isomorphism
$$\vV_0(v,\Dd(d,d'))\cong \Dd_0(v\odot d,d')$$
As an important remark, if $\vV$ can be considered as a $\vV$-category (meaning that it is monoidal closed) then
it has powers given by $v\pitchfork w = [v,w]$ and copowers given by $v \odot w = v \otimes w$.

This Kelly \cite[Lemma 3.68]{kellyBasicConceptsEnriched1982} gives us a way to explicitly compute pointwise Kan extensions.
\begin{lemma}\amslabel{lem:end-existence}
  Let $\Cc$ and $\Dd$ be $\vV$-categories for $\vV$ a symmetric monoidal closed category
  and let $P:\Cc\op\otimes\Cc\xrightarrow{}\Dd$ be a $\Dd$-valued $\vV$-distributor on $\Cc$ (i.e.\ a $\vV$-functor).

  If the following powers of $\Dd$, conical products and conical equalizer exist,
  then the end of $P$ exists, and is the equalizer of the diagram:
  \begin{center}
    \begin{tikzcd}[ampersand replacement=\&]
      {} \& {} \& {\prod_{c\in\Cc_0}P(c,c)} \&\&\& {\prod_{(c,c')\in(\Cc_0)^2}\Cc(c,c')\pitchfork P(c,c')}
      \arrow["{\phi}", shift left=2, from=1-3, to=1-6]
      \arrow["{\psi}"', shift right=2, from=1-3, to=1-6]
    \end{tikzcd}
  \end{center}
  where the top arrow (of $\Dd_0$) is defined, at component $(c,c')\in(\Cc_0)^2$, as the composite
  $$\prod_{c\in\Cc_0}P(c,c)\xrightarrow{\pi_c}P(c,c)\xrightarrow{(P(c,-)_{c,c'})^\dashv}\Cc(c,c')\pitchfork P(c,c')$$
  where $(P(c,-)_{c,c'})^\dashv$ is the adjunct of the following morphism of $\vV_0$:
  $$\Cc(c,c')\cong^{\lambda^{-1}}I\otimes\Cc(c,c')\xrightarrow{\id^\Cc_c\otimes\Cc(c,c')}\Cc(c,c)\otimes\Cc(c,c')
  \xrightarrow{P_{(c,c),(c,c')}}\Dd(P(c,c),P(c,c'))$$
  that we denote $P(c,-)_{c,c'}$. The bottom arrow of $\Dd_0$ is given at component $(c,c')\in(\Cc_0)^2$ by the adjunct
  $$P(c',c')\xrightarrow{(P(-,c')_{c,c'})^\dashv}\Cc(c,c')\pitchfork P(c,c')$$
  of the morphism of $\vV_0$
  $$\Cc(c,c')\cong^{\rho^{-1}}\Cc(c,c')\otimes I\xrightarrow{\Cc(c,c')\otimes\id^\Cc_{c'}}\Cc(c,c')\otimes\Cc(c',c')
  \xrightarrow{P_{(c',c'),(c,c')}}\Dd(P(c',c'),P(c,c'))$$
  that we denote $P(-,c')_{c,c'}$.

  Dually, if the following copowers of $\Dd$, (conical) coproducts and (conical) coequalizer exist,
  then the coend of $P$ exists, and is the coequalizer of the diagram:
  \begin{center}
    \begin{tikzcd}[ampersand replacement=\&, outer sep=3pt]
      {} \& {} \& {\sum_{c\in\Cc_0}P(c,c)} \&\&\& {\sum_{(c,c')\in(\Cc_0)^2}\Cc(c,c')\odot P(c,c')}
      \arrow["{(\sum_{c\in\Cc_0}P(c,c)\xleftarrow{\i_c}P(c,c)\xleftarrow{(P(c,-)_{c,c'})^\vdash}\Cc(c,c')\odot P(c,c'))_{(c,c')\in(\Cc_0)^2}}"', shift right=2, from=1-6, to=1-3]
      \arrow["{(\sum_{c\in\Cc_0}P(c,c)\xleftarrow[\i_{c'}]{}P(c',c')\xleftarrow[(P(-,c')_{c,c'})^\vdash]{}\Cc(c,c')\odot P(c,c'))_{(c,c')\in(\Cc_0)^2}}", shift left=2, from=1-6, to=1-3]
    \end{tikzcd}
  \end{center}
  where this time, the adjuncts of the morphisms $P(c,-)_{c,c'}$ and $P(-,c')_{c,c'}$ of $\Dd_0$ have to be understood with respect
  to the copowers.
\end{lemma}

\begin{proof}
  Sketch of proof: the formulas can be shown to be true for $\Dd=\vV$ using the definition of an end.
  Now for $\Dd$-valued distributors, $e\in\Dd_0$ is the end of $P$ if there is an isomorphism
  $$\Dd(d,\int_{c:\Cc}P(c,c))\cong\int_{c:\Cc}\Dd(d,P(c,c))$$
  natural in $d$. Now
  \begin{align*}
    \Dd(d,\Eq(\phi,\psi))
    &\cong \Eq(\Dd(d,\phi),\Dd(d,\psi))
  \end{align*}
  where the equalizer on the right is the one of the lemma for the $\vV$-distributor $\Dd(d,P(-,=))$.
  Indeed, by continuity of $\Dd(d,-)$,
  $$\prod_{c\in\Cc_0}\Dd(d,P(c,c))\xrightarrow{\Dd(d,\phi)}
  \prod_{c,c'\in\Cc_0}\Dd(d,\Cc(c,c')\pitchfork P(c,c'))$$
  is given at component $(c,c')$ by
  $$\prod_{c\in\Cc_0}\Dd(d,P(c,c))\xrightarrow{\pi_c}\Dd(d,P(c,c))
  \xrightarrow{\Dd(d,(P(c,-)_{c,c'})^\dashv)}\Dd(\Cc(c,c')\pitchfork P(c,c'))$$
  which because of powering is in fact
  $$\prod_{c\in\Cc_0}\Dd(d,P(c,c))\xrightarrow{\pi_c}\Dd(d,P(c,c))
  \xrightarrow{\Dd(d,(P(c,-)_{c,c'}))^\dashv}\Dd(\Cc(c,c')\pitchfork P(c,c'))$$
  therefore $\Eq(\Dd(d,\phi),\Dd(d,\psi))$ is the end of $\Dd(d,P(-,=))$.
\end{proof}

\begin{lemma}\amslabel{lem:pt-kan-are-loc-kan}
  Consider a span of $\vV$-functors $\Cc'\xleftarrow{H}\Cc\xrightarrow{F}\Dd$.
  If the pointwise left (resp. right) Kan extension $(\Lan_H F,\lambda:F\xRightarrow{}H^\ast\Lan_H F)$ (resp. $(\Ran_H F,\rho:H^\ast\Ran_H F\xRightarrow{}F)$) exists
  then we have a $\vV$-natural isomorphism (in $G$)
  $$[\Lan_H F,G]_\vV\cong[F,HG]_\vV$$
  and on the unenriched side, the unenriched natural isomorphism is given by
  $$(\alpha : \Lan_H F \xRightarrow{} G) \xmapsto{} (F\xRightarrow{\lambda}H\Lan_H F\xRightarrow{H\ast\alpha}HG)$$
  Respectively, for the pointwise right Kan extension,
  $$[G,R]_\vV\cong[HG,F]_\vV$$
\end{lemma}

\begin{proof}
  We do it for the left Kan extension, the same arguments apply dually for the right Kan extension.
  \begin{align*}
    [\Lan_H F,G]_\vV
    &\cong [\int^{c:\Cc}\Cc'(H(c),-)\odot F(c),G]_\vV \quad \text{by definition of L} \\
    &\cong \int_{c:\Cc}[\Cc'(H(c),-)\odot F(c),G]_\vV \quad \text{by cocontinuity of the }\hom \\
    &\cong \int_{c:\Cc}\int_{c'\in\Cc'}\Dd(\Cc'(H(c),c')\odot F(c),G(c')) \quad \text{by definition of $\vV\Cat$ as a $\vV$-category} \\
    &\cong \int_{c:\Cc}\int_{c'\in\Cc'}[\Cc'(H(c),c'),\Dd(F(c),G(c'))] \quad \text{by definition of the copower} \\
    &\cong \int_{c:\Cc}\Dd(F(c),G(H(c))) \quad \text{by ninja Yoneda lemma on } \Dd(F(c),G(-)) \\
    &\cong [F,HG]_\vV\quad \text{by definition}\\
  \end{align*}
\end{proof}

This following crucial lemma shows that in the case we consider a Kan extension along a full subcategory inclusion, then the Kan extension is a ``real''
extension.

\begin{lemma}\amslabel{lem:enriched-kan-ext-along-fully-faithful}
  Consider a span of $\vV$-functors $\Cc'\xhookleftarrow{H}\Cc\xrightarrow{F}\Dd$, such that $H$ is fully faithful\footnote{recall that is the enriched sense, it means that $H_{c,d}:\Cc(c,d)\xrightarrow{}\Cc'(H(c),H(d))$
    are isomorphisms in $\vV$ for all pairs of objects $(c,d)$ of $\Cc$.}.
  If the pointwise left (resp. right) Kan extension $\lambda:F\xRightarrow{}HL$ (resp. $\rho:HR\xRightarrow{}F$) exists,
  then $\lambda$ (resp. $\rho$) is in fact an isomorphism.
\end{lemma}

\begin{proof} 
  We give a sketch of proof based on the proof of Kelly \cite[Proposition 4.23]{kellyBasicConceptsEnriched1982}, but for the right Kan extension,
  and using only ends.

  First consider the Yoneda $\vV$-functor $\Cc(-_1,-_2):\Cc\op\xrightarrow{}[\Cc,\vV]_\vV$, and the composite $\vV$-functor
  $\Cc'(H(-_1),H(-_2)):\Cc\op\xrightarrow{}[\Cc,\vV]_\vV$. Then
  $$H_{-_1,-_2}:\Cc(-_1,-_2)\xRightarrow{}\Cc'(H(-_1),H(-_2))$$
  defined componentwise by
  $$H_{c,c'}:\Cc(c,c')\xrightarrow{}\Cc'(H(c),H(c'))$$
  is $\vV$-natural, and is an isomorphism if and only if $H$ is fully faithful, because by definition $H$ is fully faithful
  if and only if $H_{c,c'}:\Cc(c,c')\xrightarrow{}\Cc'(H(c),H(c'))$ is an isomorphism for all $(c,c')\in(\Cc_0)^2$.

  Now by Yoneda's lemma, there is a $\vV$-natural isomorphism
  $$\int_{c:\Cc}\Cc(-,c)\pitchfork F(c)\xRightarrow{y} F(-)$$
  and one may check that the counit $\rho$ of the right Kan extension is in fact
  \begin{center}
    \begin{tikzcd}[ampersand replacement=\&]
      {(H\Ran_HF)(c)=\int_{c':\Cc}\Cc'(H(c),H(c'))\pitchfork F(c')} \&\& {F(c)} \\
      {\int_{c':\Cc}\Cc(c,c')\pitchfork F(c')}
      \arrow["{\rho_c}", from=1-1, to=1-3]
      \arrow["{\int_{c':\Cc}H_{c,c'}\pitchfork F(c')}"', from=1-1, to=2-1]
      \arrow["{y_c}"', from=2-1, to=1-3]
    \end{tikzcd}
  \end{center}
  Now because $y$ is a $\vV$-natural isomorphism, if $H$ is fully faithful, then $H_{-_1,-_2}$ is a $\vV$-natural isomorphism and
  therefore so is $\int_{c':\Cc}H_{c,c'}\pitchfork F(c')$ and finally so is $\rho$.
\end{proof}

\section{Free $\eE$-categories for a bicomplete elementary topos $\eE$}
\label{sec:free-e-categories}

\begin{defi}
  Let $\vV=(\vV_0,\otimes, I)$ be a closed symmetric monoidal category, a (small) $\vV$-quiver $Q=(Q_0,Q(-,=))$
  is given by a set \emph{of vertices} $Q_0$ and for all $(x,y)\in Q^2$, an object $Q(x,y)$ \emph{of edges} of $\vV_0$.
  
  A \emph{$\vV$-quiver morphism} $f:Q\xrightarrow{} R$ is a collection $(f,(f_{x,y})_{(x,y)\in Q^2})$ where $f$ is
  a function from $Q_0$ to $R_0$ and $f_{x,y}$ a morphism of $\vV_0$ from $Q(x,y)$ to $R(f(x),f(y))$.
\end{defi}

\begin{defi}
  Let $\vV=(\vV_0,\otimes, I)$ be a symmetric closed monoidal category, we say $\vV$ admits free $\vV$-categories if the \emph{forgetful} (unenriched) functor $$\vV\Cat \ni \Cc \xmapsto{} (\Cc_0,\Cc(-,=)) \in \vV\Quiv$$ has a left adjoint.
\end{defi}


\begin{prop}
  A bicomplete elementary topos $\eE$ admits free $\eE$-categories.
\end{prop}

\begin{proof}
  The free $\eE$-category $\Cc$ generated by the $\eE$-quiver $Q$ has for set of objects $Q_0$ and for all $x,y\in Q_0$, $$\Cc(x,y)=\sum_{n\in \nN}\sum_{x_1,x_2,\dots,x_{n-1}\in V_0}Q_0(x,x_1)\times Q_0(x_1,x_2)\times Q_0(x_2,x_3)\times \cdots \times Q_0(x_{n-1},y)$$ so that the identity of $x$ is given by the coproduct injection (for $n=0$, and path $(x)$) $\terminal \xhookrightarrow{}Q(x,x)$ and composition works because sums distribute over products in this setting (a topos being an extensive category).
  
  Now the adjunction itself. Let $f$ be an $\eE$-quivers morphism between a quiver $Q$ and an $\eE$-category $\Dd$ seen as an $\eE$-quiver, and denote $\Cc$ the free $\eE$-category generated by $Q$. Then the adjunct of $f$ is an $\eE$-functor $F$ such that $F_0=f_0:\Cc_0\xrightarrow{}\Dd_0$ where $\Cc_0=Q_0$ by definition. Now for all $x,y\in\cC_0$, $$\Cc(x,y)=\sum_{n\in \nN}\sum_{x_1,x_2,\dots,x_{n-1}\in \cC_0}Q(x,x_1)\times Q(x_1,x_2)\times Q(x_2,x_3)\times \cdots \times Q(x_{n-1},y)$$ so $F_{x,y}:\Cc(x,y)\xrightarrow{}\Dd(F_0(x),F_0(y))$ is defined using the universal property of the coproduct: for all $n\in\nN, x_1,x_2,\dots,x_{n-1}\in\cC_0$, $\restr{F_{x,y}}{\prod_{i=0}^{n-1}Q(x_i,x_{i-1})}$ is defined by
  \begin{center}
\[\begin{tikzcd}
	{\prod_{i=0}^{n-1}Q(x_i,x_{i-1})} && {\dD(F_0(x),F_0(y))} \\
	& {\prod_{i=0}^{n-1}\dD(F_0(x_i),F_0(x_{i-1}))}
	\arrow["{\restr{F_{x,y}}{\prod_{i=0}^{n-1}Q(x_i,x_{i-1})}}", from=1-1, to=1-3]
	\arrow["{\prod_{i=0}^{n-1}f_{x_i,x_{i-1}}}"', from=1-1, to=2-2]
	\arrow["{\comp_{F_0(x),F_0(x_1),\dots,F_0(x_{n-1}),F_0(y)}}"', from=2-2, to=1-3]
      \end{tikzcd}\]
  \end{center}
  where $\comp_{x,x_1,x_2,\dots,x_{n-1},y}$ is the iteration of composition of the $\eE$-category $\Dd$ (it can be defined in different manners thanks to associativity), with the convention that for $n=0$, $\comp_y:\terminal\xrightarrow{}\Dd(y,y)$ is the identity of $y$. The newly defined $F$ preserves identities because by definition, the identity of $x\in\Cc_0$ is the global element corresponding to the coproduct injection for $n=0,x=x$. $F$ preserves it because of the convention that $\comp_{F_0(x)}$ is the identity of $F_0(x)$. It preserves composition of morphisms by definition of composition in the free category $\Cc$.
\end{proof}

\bibliographystyle{plain}
\bibliography{main}
\end{document}